\documentclass[aps,onecolumn, floatfix, notitlepage, pre]{revtex4-1}
\usepackage{amsmath}
\usepackage{amsfonts}
\usepackage{amssymb}
\usepackage{graphicx}
\usepackage{bm}
\usepackage{color}
\usepackage{ulem}
 
 \usepackage{epsfig}
\usepackage{graphicx}
\usepackage{chngcntr}
\counterwithout{figure}{section}
\usepackage{rotating}
\usepackage{amssymb}
\usepackage{amsmath}
\usepackage{physics}
\usepackage{multirow}
\usepackage{float}
\usepackage{caption, subcaption}
\usepackage{mathtools}
\makeatletter
\renewcommand\paragraph{\@startsection{paragraph}{4}{\z@}%
            {-2.5ex\@plus -1ex \@minus -.25ex}%
            {1.25ex \@plus .25ex}%
            {\normalfont\normalsize\bfseries}}
\makeatother
\begin{document} 

\title{\bf Phase transitions in the Blume-Capel model with trimodal and Gaussian random fields}

\author{Soheli Mukherjee,}
\email{soheli.mukherjee@niser.ac.in}
\author{Sumedha}
\email{sumedha@niser.ac.in}
\affiliation{School of Physical Sciences, National Institute of Science Education and Research, Jatni - 752050,India}
\affiliation{Homi Bhabha National Institute, Training School Complex, Anushakti Nagar, Mumbai 400094, India}

\begin{abstract}

 We study the effect of different symmetric random field distributions: trimodal and Gaussian on the phase diagram of the infinite range Blume-Capel model.  For the trimodal random field, the model has a very rich phase diagram. We find   three new ordered phases, multicritical points like tricritical point (TCP), bicritical end point (BEP), critical end point (CEP) along with some multi-phase coexistence points. We also find re-entrance at low temperatures for some values of the parameters. On the other hand  for  the Gaussian distribution the phase diagram consists of a continuous line of transition followed by a first order transition line, meeting at a TCP. The TCP vanishes for higher strength of the random field. In contrast to the trimodal case, in Gaussian case no new phase emerges.

\end{abstract}

\maketitle

\section{Introduction}

Spin systems with random field disorder are an important class of models studied extensively as prototypes for collective phenomenon in systems with quenched disorder \cite{experiments, experiments2, experiments3, wetting, binary, interface, alloy, binary2, network}. They model diluted antiferromagnets like $Fe_x Zn_{1-x} F_2, \,\, Rb_2 Co_x Mg_{1-x} F_4$ in the presence of a uniform magnetic field \cite{experiments, experiments2, experiments3}. In addition to this, many random systems like  prewetting transition on a disordered substrate \cite{wetting}, binary fluid mixtures in random porous media \cite{binary}, phase transitions and interfaces in random media \cite{interface}, structural phase transitions in random alloys \cite{alloy}, binary fluids in gels \cite{binary2}, collective effects  induced by imitation and social pressure on society via network models \cite{network} are modelled by ferromagnets in the presence of random field.  Geophysical models of marine climate pattern \cite{geo-1}, identification of subsurface soil patterns \cite{geo-2}, analysis of molecular structures \cite{chem}, biomedical imaging \cite{biomed-1, biomed-2}, population genetics \cite{gene}, data science \cite{data} and many more problems in other disciplines \cite{reference} have also been modelled using random fields.

 Phase diagrams of the random field systems are known to depend  non-trivially on the distribution of the random field \cite{aharony, pytte, galam-dist, andelman}. In the case of infinite range random field Ising model (RFIM) for the Gaussian distribution of the random field the phase diagram has a line of continuous transitions separating the ordered and disordered phases at all strengths of the random field \cite{pytte}. On the other hand, for the bimodal distribution of the random field, the transition changes to a first order transition as the random field strength increases \cite{aharony}. In three dimensions the nature of the transition is still debated \cite{fytas2, fytas}. Most recent numerical studies in three dimensions find the transition to be continuous independent of the nature of the random field distribution \cite{fytas2, fytas}. Other distributions like  trimodal \cite{trimodal1, trimodal2, trimodal3,trimodal4, numerical2}, double Gaussian \cite{doublegaussian}, triple Gaussian \cite{trip-gauss}, asymmetric trimodal \cite{asytrimod}, asymmetric bimodal distribution \cite{asybimod} have also been studied for RFIM.  In all these studies the nature of the phase diagrams depends on the symmetries of the random field distribution.

The ferromagnetic spin-1 system with crystal field known as the Blume-Capel model is an important spin model that models many physical systems like  multicomponent fluid mixtures \cite{fluid-mixture}, $^3{He} - ^4{He}$ mixtures \cite{beg}, binary alloys \cite{alloy2}, metamagnets \cite{metamag, tcp}, inverse melting and inverse freezing \cite{reentrance, reentrance1}. In the pure case, its phase diagram consists of a continuous, and a first-order transition line. These two transition lines meet at a tricritical point (TCP). This model was first studied by Blume \cite{blume} and Capel \cite{capel} in order to explain the first order transition in the $UO_2$.

 Blume-Capel model in the presence of bimodal random field distribution has been studied earlier in \cite{rfbc, santos, albayrak2, sspin, bc-randomnetwork}. In \cite{rfbc}, using the mean-field method, the phase diagram in the temperature ($T$)- random magnetic field strength ($h$) plane was determined for different values of the crystal field ($\Delta$). They identified five different $T-h$ phase diagrams \cite{santos}. The model was revisited to obtain the $T-\Delta$ phase diagrams at fixed values of $h$. It was shown that the projection of the phase diagrams are similar in the both $T-\Delta$ and $T-h$ planes. The RFBC model has also been studied on the Bethe lattice  for bimodal \cite{albayrak2} and the equal  trimodal distribution \cite{albayrak}. Here, the presence of two first order transition lines along with a second order transition line was reported. The RFBC model for bimodal random field distribution was also studied in a random network with finite connectivity  using the replica trick \cite{bc-randomnetwork}.

In this paper, we study the trimodal distribution of the random field on a fully connected graph where there is a quenched random field at each site chosen from the distribution
\begin{equation}\label{dist tri}
p(h_i)= p \delta(h_i)+ \frac{1-p}{2} [ \delta(h_i+h) + \delta(h_i-h)]
\end{equation}
We study all values of $0 \leq p \leq 1$. For $p=0$ the distribution is the same as the bimodal distribution and for $p=\frac{1}{3}$ it is the equal trimodal distribution. We also study the Gaussian distribution of the random field.  Trimodal RFIM is relevant to study the diluted antiferromagnets in a uniform field, where the field conjugate to the antiferromagnetic order parameter takes three values \cite{experiments3}. In trimodal distribution a fraction of $p$ spins are free from the external magnetic field. This feature is similar to the behaviour of the Gaussian distribution with maxima at zero field. We solve the model using the method based on large deviation theory (LDT) \cite{dembo, ldp} which has been used recently to solve random field problems with discrete \cite{disc-ldt} and continuous spins \cite{cont-ldt}.

In the case of trimodal random field distribution, we find that the three new ordered phases emerge at $T=0$. The model has $4$ ordered and $2$ disordered phases that are separated by  first order transition lines. We classify the $T=0$ phase diagram into five categories depending on the value of $p$ : $p=1$, $\frac{1}{3} < p < 1$, $p=\frac{1}{3}$, $0 <p< \frac{1}{3}$ and $p=0$.

 For finite $T$  with trimodal random field distribution, we find that the phase diagrams can again be divided into  five categories that depend on the value of $p$ just like the $T=0$ phase diagrams. We have studied the phase diagram in the $T-\Delta$ and $T-h$ planes. For $p=0$, there are six different $T-\Delta$ and seven different $T-h$ phase diagrams. For $0 < p < \frac{1}{3}$ there are eight and seven different phase diagrams in the $T-\Delta$ and $T-h$ planes respectively. At $p = \frac{1}{3}$, there are nine different  $T-\Delta$ and eight different $T-h$ phase diagrams. For $\frac{1}{3} < p < 1$, there are eight and nine different phase diagrams in the $T-\Delta$ and $T-h$ planes respectively.  For $p=0$ (the bimodal distribution), we find three additional  phase diagrams which were not reported earlier in \cite{rfbc, santos}. We find that the RFBC model exhibits re-entrance  in the $T-h$ plane for some values of $\Delta$ for the equal ($p = \frac{1}{3}$) trimodal distribution.

We also studied the RFBC model for Gaussian distribution. The strength of disorder is measured by the value of $\sigma^2$, the variance of the Gaussian distribution. At $T=0$, $\sigma$ behaves like a temperature and $\Delta-\sigma$ phase diagram is similar to the $\Delta-T$ phase diagram of the pure Blume-Capel model. We observe that in contrast to the trimodal case, there are no new ordered phases at high values of the disorder strength.

At finite $T$ in presence of Gaussian distribution we find that for weak disorder, the phase diagram consists of a second order transition followed by a first order transition that meets at a TCP for weak disorder. The TCP moves towards lower temperature as $\sigma$ increases and above a critical value $\sigma_c$, the TCP vanishes, and the phase diagram  consists only of a second order line. The RFBC model for Gaussian distribution with infinite range interaction was studied earlier using the effective field theory \cite{sspin}. They reported only a continuous transition line in the presence of disorder for all temperatures. We show that first order transition vanishes only for $\sigma > \sigma_c$.

 For the  random field ferromagnetic $O(n)$ models Aharony conjectured that the phase transition at low temperature is first order (second-order) if the random field distribution function is symmetric and has a minimum (maximum) at zero field \cite{aharony}. Later this criterion was refined in \cite{galam-dist, andelman} based on the maxima of the distribution function.  Trimodal distribution has been argued  to  be a good approximation of the Gaussian distribution for $p=\frac{1}{3}$ in \cite{trimodal1}. In \cite{trimodal2, trimodal3, trimodal4, numerical2}, for RFIM the trimodal and the Gaussian distributions were found to have similar phase diagram with only continuous transition at all temperatures. Recently,  for random field XY model (RFXY) model the equation of the line of continuous transition was shown to be  same for all symmetric distributions on a fully connected graph \cite{cont-ldt}. We find rather surprisingly, no similarity in the phase diagrams of the symmetric trimodal $p \geq \frac{1}{3}$ and Gaussian distribution for the RFBC.

The phase diagrams for the trimodal RFBC model consist of multicritical points like bicritical end point (BEP), critical end point (CEP), TCP and some multiple phase coexistence points like $A_5$, $A_6$ and $A_7$. TCP and BEP are the multicritical points at which three and two critical lines end respectively. BEP has been reported as an ordinary critical point earlier in the studies of bimodal random field RFBC model \cite{rfbc}. CEP is a critical point where a line of second order transition terminates on a first order transition line. $A_n$ points are defined as the coexistence point of $n$ phases. We found that the multicritical points as well as the $A_n$ points appear multiple times in the phase diagram. In order to understand the origin of the multicritical points like TCP and BEP  we study their coordinates in the $\Delta-h$ plane. We observe that the TCP and the BEP coordinates in the $\Delta-h$ plane closely follow the phase boundaries of the $T=0$ phase diagram. We hence conclude that the location of TCPs and BEPs strongly depend on $p$, $\Delta$ and $h$, and weakly on $T$.

The present work is organized as follows : In Sec. II we give the solution of the random field Blume-Capel model on a fully connected graph. In Sec. III we study the ground state phases and the ground state phase diagram for both the distributions. In Sec. IV, we determine the finite temperature phase diagrams of both the distributions. In Sec. V
we look at the projections of the TCPs and BEPs in the case of trimodal distribution on the $\Delta-h$ plane. In Sec. VI, we summarize and discuss the main results.
\section{Model and Formulation}\label{sec1}

The Hamiltonian for the RFBC model with $N$ spins on a fully connected graph is given by 

\begin{equation}\label{eq1}
\mathcal{H}= -\frac{1}{2 N} (\sum_i s_i)^2 + \Delta \sum_i s_i ^2 - \sum_i (h_i+H) s_i
\end{equation}
here $s_i$ takes values $\pm 1, 0$; $\Delta$ is the crystal field of the system which we take to be $ \geq 0$ favoring $s=0$ spins; $H$ is the uniform external magnetic field and $h_i$ is the quenched local random field drawn from a distribution $P(h)$.

We study two distributions of $P(h)$ :

\textbf{(a)} the trimodal distribution  as defined in Eq. \ref{dist tri}. In this case the random field takes three values : $\pm h$ and $0$ with probabilities $\frac{1-p}{2}$ and $p$ respectively. The mean of the random field $<h_i>=0$ and the  variance $<h_i^2> - <h_i>^2 = (1-p)h^2$.

  \textbf{(b)} the Gaussian distribution defined as
\begin{equation}\label{gaussian}
		P(h_i)= \frac{1}{\sqrt{2 \pi \sigma^2 }}\,\,\,  e^{\frac{- h_i^2}{2 \sigma ^2 }}
	\end{equation}
here $\sigma^2$ is the variance of the Gaussian distribution. The mean of the distribution is zero.

 We solve the model using large deviation theory (LDT) \cite{dembo, ldp}. The probability of a spin configuration $C_N$ with magnetization $x_1 =\frac{\sum_i s_i}{N}$ and quadrupole moment $x_2 =\frac{\sum_i s_i^2}{N}$ satisfies the large deviation principle (LDP)  and can be written as 
 \begin{equation}
 P \Big (C_N: \, x_1=\frac{\sum_i s_i}{N}, \,\, x_2=\frac{\sum_i s_i^2}{N} \Big) \sim e^{- N I(x_1, x_2)}
 \end{equation}
 where $I(x_1, x_2)$ is  the rate function. The rate function $I(x_1, x_2)$  can be seen as the generalized free energy functional. Minimization of $I(x_1, x_2)$ with respect to $ x_1$ and $x_2$ gives the free energy of the system. The calculation of the rate function is given in Appendix \ref{sec1b}. 

By considering the rate function only at its fixed points, we get the generalized free energy functional which is a function of $m$, the value of $x_1$ at the fixed point. This we denote by $f(m)$ and is given by

\begin{figure}
\centering
     \begin{subfigure}[b]{0.3\textwidth}
         \centering
         \includegraphics[width=\textwidth]{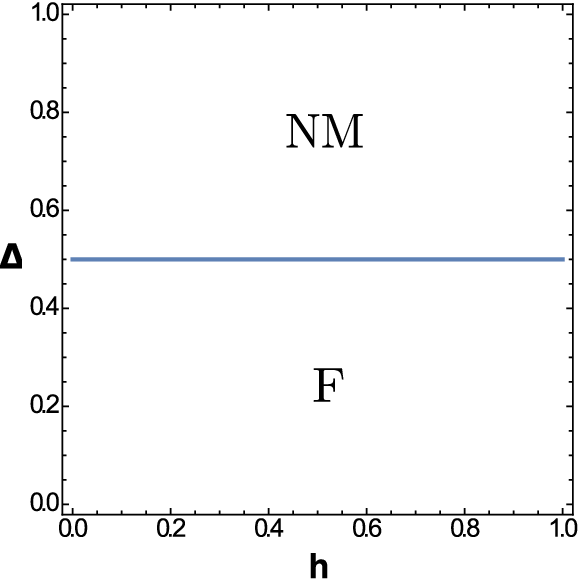}
         \caption{}
         \label{fig0}
     \end{subfigure}
     \hfill
     \centering
     \begin{subfigure}[b]{0.3\textwidth}
         \centering
         \includegraphics[width=\textwidth]{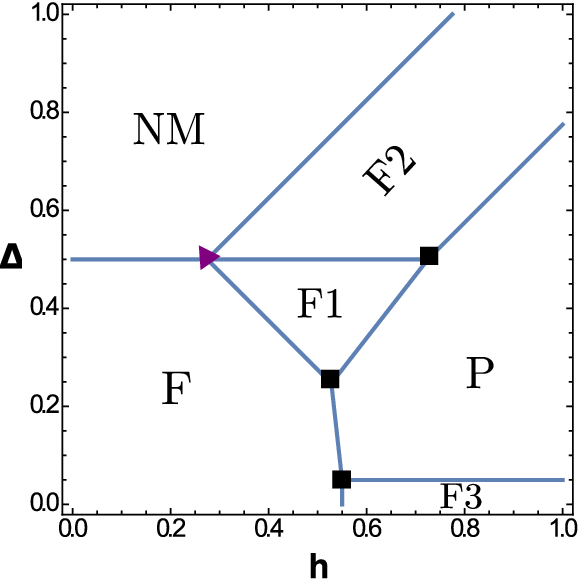}
         \caption{}
         \label{fig21}
     \end{subfigure}
     \hfill
     \centering
     \begin{subfigure}[b]{0.3\textwidth}
         \centering
         \includegraphics[width=\textwidth]{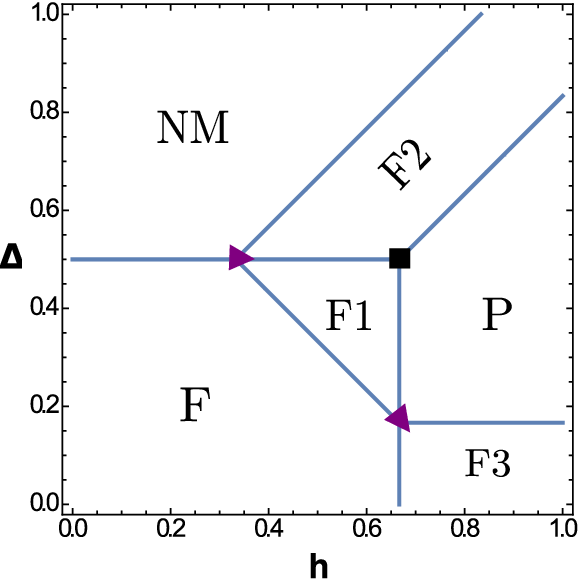}
          \caption{}
         \label{fig22}
     \end{subfigure}
     \hfill
     \begin{subfigure}[b]{0.3\textwidth}
         \centering
         \includegraphics[width=\textwidth]{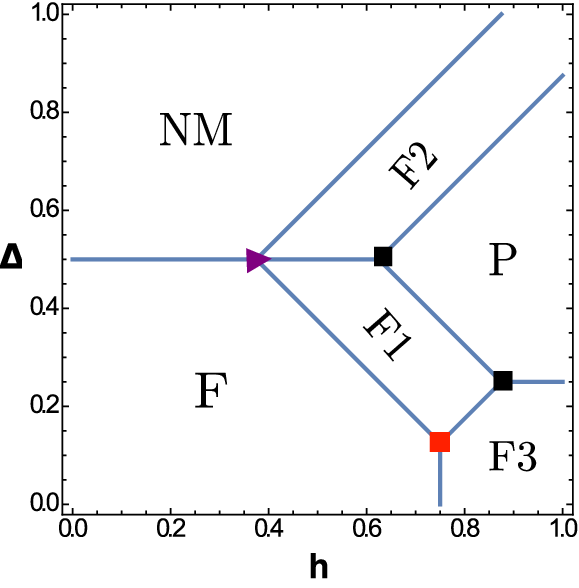}
        \caption{}
         \label{fig23}
     \end{subfigure}
     \hfill
     \centering
     \begin{subfigure}[b]{0.3\textwidth}
         \centering
         \includegraphics[width=\textwidth]{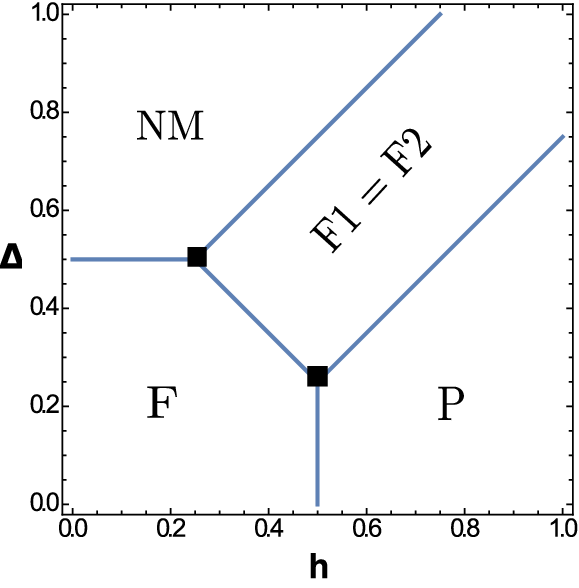}
         \caption{}
         \label{fig1}
     \end{subfigure}
     \hfill
        \caption{Ground state phase diagram for the trimodal distribution. Solid lines are the lines of first order transition. \textbf{(a)} For the pure Blume-Capel model ($p=1$). In this case there is a first order transition from the $\textbf{F} \, (m=q=1)$ to the $\textbf{NM} \, (m=q=0)$ phase at $\Delta=0.5$.  \textbf{(b)}, \textbf{(c)} and \textbf{(d)} are the phase diagram for $0 < p <1$. Each of the phase diagram contains six phases: \textbf{F}, \textbf{ F1} \, $(m=q=\frac{1+p}{2})$,  \textbf{F2} $ (m=q=\frac{1-p}{2})$, \, \textbf{F3} $(m=p, \,\, q=1)$,  \textbf{P} $(m=0, \,\, q=1-p)$, and,  \textbf{NM}. All the phases are separated by first order transition lines. There are also multi-phase coexistence points in the phase diagrams.  The purple triangles denote the $A_7$ points, black squares are the $A_5$ points and the red squares are the $A_6$ points.  Fig.\textbf{(b)} shows the phase diagram for $p=\frac{1}{10}$. This qualitatively holds for all $0< p < \frac{1}{3}$. Fig.\textbf{(c)} shows the phase diagram for  $p=\frac{1}{3}$ and Fig.\textbf{(d)} shows the phase diagram for $p=\frac{1}{2}$, this qualitatively holds for all $\frac{1}{3} < p < 1$. Fig. \textbf{(e)} is the phase diagram for $p=0$, the bimodal random field distribution. In this case, the phases \textbf{F1} and \textbf{F2} are the same, shown as \textbf{F1=F2}. }
        \label{fig2}
\end{figure}

\begin{eqnarray}\label{eqn2}
		f(m) =  \frac{\beta m^2}{2}- \,\, \Bigg < \log (1+2 e^{-\beta \bigtriangleup} \cosh { \beta (m + H +  h_i)} \,\, )\Bigg > \nonumber \\
	\end{eqnarray}
here $\langle \rangle$ is the average over the random field distribution and $m$ is the absolute magnetization. By equating the derivative of $f(m)$ with respect to $m$ to $0$, we get the equation for magnetization, which is a order parameter of the system as

\begin{eqnarray}\label{eqn1}
		m =  \Bigg< \frac{ 2 e^{ -\beta \Delta} \sinh \beta (h_i+ H +m)}{1+ 2 e^{ -\beta \Delta} \cosh \beta (h_i+ H + m)}\Bigg > \nonumber \\
	\end{eqnarray}
Another order parameter is the quadrupole moment ($q$). It is the expectation value of $s_i^2$, given by $q \,= \,\, \frac{1}{\beta} \frac{\partial{f(m)}}{\partial{\Delta}}$. We get 	

	\begin{eqnarray}\label{eqnq}
		q = \Bigg< \frac{ 2 e^{ -\beta \Delta} \cosh \beta (h_i+ H +m)}{1+ 2 e^{ -\beta \Delta} \cosh \beta (h_i+ H + m)}\Bigg > \nonumber \\
	\end{eqnarray}

\section{Zero temperature phase diagram}\label{sec2}

We first determine the $T=0$ phase diagram in the $\Delta - h$ plane for both the distributions in this section.

\subsection{Trimodal distribution}\label{sec2a}

Using the $P(h_i)$ from Eq. \ref{dist tri} in Eq. \ref{eqn2}, the free energy functional of the system at $H=0$ is

\begin{eqnarray}\label{eq2}
f(m) &=& \frac{\beta m^2}{2} -p \log \Bigg( 1+ 2 e^{-\beta \Delta} \cosh \beta m\Bigg) 
 -  \frac{1-p}{2} \log \Bigg (1+ 2 e^{-\beta \Delta} \cosh \beta (-h+m)\Bigg) \nonumber \\
&-&  \frac{1-p}{2}  \log \Bigg( 1+ 2 e^{-\beta \Delta} \cosh \beta (h+m)\Bigg) 
\end{eqnarray}
For $\beta \rightarrow \infty$, the ground state rate function  given by, $\Phi(m)= \lim \limits_{\beta \rightarrow \infty} \frac{1}{\beta}f(m) $  is

\begin{eqnarray}
\Phi(m) &=& \frac{m^2}{2} - p \,\, \mid m - \Delta \mid \,\,  \Theta(m - \Delta ) 
  -  \frac{1-p}{2} \,\,  \, \mid m + h - \Delta \mid \,\,  \Theta(m +h  - \Delta ) \nonumber \\
 & - & \frac{1-p}{2} \,\,  \mid h - \Delta - m \mid \,\,  \Theta(h  - \Delta - m )   -   \frac{1-p}{2} \,\, \mid m -h - \Delta \mid \,\,  \Theta(m- h  - \Delta  )  \nonumber \\
\end{eqnarray}
here $\Theta(x)$ is the Heaviside step function with $\Theta(x)=1$ for $x > 0$, and $\Theta(x)=0$ for $x \leq  0$. The disorder averaged ground state energy is $E =$  $\min \limits_m \Phi(m)$.

  Depending on the values of the parameters $\Delta$, $h$ and $p$ there are six phases that exist for any $0 < p< 1$. There are  four  ferromagnetic phases, one paramagnetic and one non-magnetic phase.  The four ferromagnetic phases are defined as : $\textbf{F} \equiv \,\, m=q=1$, $\textbf{F1} \equiv \,\, m=q=\frac{1+p}{2}$, $\textbf{F2} \equiv \,\,m=q=\frac{1-p}{2} $ and $\textbf{F3} \equiv \,\, m=p, \,\, q=1 $. For large $\Delta$ the spins are more likely to be in $0$ state, and there is a non-magnetic phase (\textbf{NM}) with  $m=q=0$. Whereas for small $\Delta$ and large $h$ the spins tend to align along the local random fields, which gives rise to a paramagnetic phase (\textbf{P}) with $ m=0, \,\, q=1-p$.

 \begin{figure}
\centering
\includegraphics[width=0.55\textwidth]{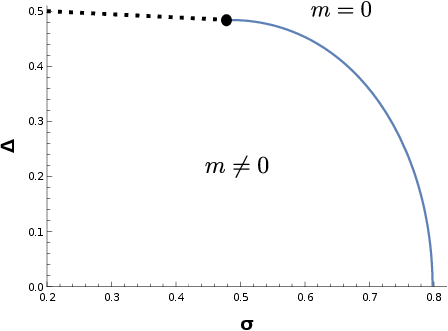}       
\caption{Ground state phase diagram for the Gaussian field distribution. Dotted line is the line of first order transitions and solid line is the line of second order transitions. Solid circle is the TCP. There is one ordered phase ($m \neq 0$) and one disordered phase ($m=0$) in the phase diagram. The transition is first order for small $\sigma$. As $\sigma$ increases, the transition changes to second order at a TCP with the coordinates $\sigma_{TCP}=\Delta_{TCP}=\sqrt{\frac{2}{e \pi}}.$ }  
   \label{fig3}
\end{figure}

 We find that the \textbf{F2} phase persists even when $\Delta$ and $h$ approach infinity. This is because the variance of the distribution increases as $h^2$. Hence, as $h$ increases, the spins prefer to be $ \pm 1$ state to take advantage of the energy lowering in a given realization due to the spin imbalance. This results in a competition between the crystal field ($\Delta$) and the random field ($h$) and we get the ferromagnetic phase \textbf{F2}  even at very large values of $\Delta$ and $h$.

For the special case $p=0$ (bimodal), only one new ferromagnetic phase, which we call \textbf{F1=F2}  phase ($m=q=\frac{1}{2}$) emerges. Whereas for $0 < p<1$, there are three new ferromagnetic phases (\textbf{F1},  \textbf{F2}, \textbf{F3}). Depending on the value of $p$ we find  that there are five different phase diagrams  (see Fig.\ref{fig2}). Below we describe these phase diagrams.


\subsubsection{$p = 1  :$ Pure Blume-Capel model}

In the pure Blume-Capel model, there is one ordered phase (\textbf{F}) and one disorder phase (\textbf{NM}). There is a line of first order transition between these two phases at $\Delta= \frac{1}{2}$ for all values of $h$, as shown in Fig. \ref{fig0}.

\subsubsection{$0 < p < 1$} \label{subsec3}

For any $0 < p <1$, there are always six phases in the system. All the phases are separated by first order transition lines. For all $0 < p <1$ there are six first order phase boundaries that are always present. These are :  the \textbf{F3} phase is separated from the \textbf{F} phase by a first order transition line parallel to the $\Delta$ axis at $h = \frac{1+p}{2}$, phases \textbf{F1} and \textbf{F2} and phases \textbf{F} and \textbf{NM} are separated via a first order transition line parallel to $h$ axis at $\Delta= \frac{1}{2}$,  phases \textbf{P} and \textbf{F3} are separated by a first order transition line  parallel to $h$ axis at $\Delta = \frac{p}{2}$. The phase \textbf{F2} is separated from phase \textbf{NM} via the first order transition line  $\Delta - h = \frac{1}{4} (1-p)$. The phase \textbf{F2} is separated from the phase \textbf{P} via the first order transition line $ h - \Delta  =  \frac{1}{4} (1-p)$. The  phases \textbf{F} and \textbf{F1} are separated by the first order transition line given by the solution of the equation   $ (1-p) (\Delta + h) = \frac{1}{4}(3 - p^2 - 2 p)$. Apart from these there are some other first order transition lines in the phase diagrams which depend on the range of $p$.

Apart from the first order transition lines, there are multi-phase coexistence points in the ground state phase diagram. We denote  seven-phase coexistence point  which is a coexistence of six ordered and one disordered phase by $A_7$, the six-phase coexistence point which is a coexistence of six ordered phases by $A_6$, and the five phase coexistence point which is a point of coexistence of four ordered and one disordered phase by $A_5$ (see Fig. \ref{fig2}). There are three possible phase diagrams depending on the value of $p$ : $0< p < \frac{1}{3}$, $p=\frac{1}{3}$ and $\frac{1}{3} < p < 1$.

\begin{itemize}

\item $ \mathbf{0 < p < \frac{1}{3}:}$ \,\, For the range  $0 < p < \frac{1}{3}$, there are two more first order phase boundaries apart from the  ones mentioned above. The phases \textbf{F1} and \textbf{P} are separated by the first order transition line given by the equation : $(3p -1) \Delta + (1-p) h = \frac{1}{4} (1+p)^2$. Another is the first order phase boundary between the phases \textbf{F} and \textbf{P}  given by the equation : $p \Delta +(1-p) h = \frac{1}{2}$. At the junction of these first order transition lines there are  multi-phase coexistence points. There is one $A_7$ point at the junction of the \textbf{F-NM-F1-F2} phases located at ($\Delta=\frac{1}{2}, \,\, h=\frac{1+p}{4}$), and three $A_5$ points  at the junction of \textbf{F-P-F1} phases at ($\Delta=\frac{1-p^2 - 2 p}{4 (1- 2p)}, \,\, h=\frac{p^3 + 2p^2- 5p +2}{4 (1- 2p) (1-p)}$),  \textbf{F-P-F3} phases at ($\Delta=\frac{p}{2}, \,\, h=\frac{1+p}{2}$),  and  \textbf{P-F1-F2} phases at ($\Delta=\frac{1}{2}, \,\, h=\frac{3 - p}{4}$). Fig. \ref{fig21} shows the ground state phase diagram for $p=\frac{1}{10}$. The purple triangle represents the $A_7$ point and the solid black squares represent the $A_5$ points.

\begin{figure}
\centering
\includegraphics[scale=0.7]{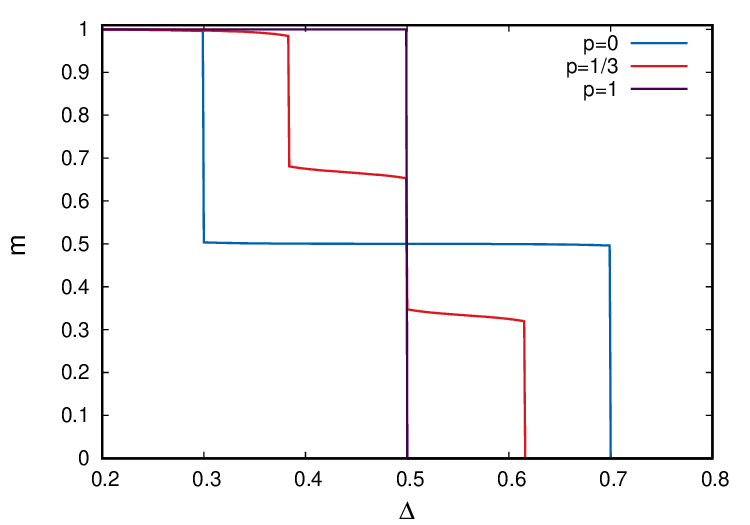}
\caption{Plot of m vs $\Delta$ for $T=0.05$ and field $h \, =0.45$ for  $p = 1, \,\, \frac{1}{3}$ and $0$. The first order transition from the \textbf{F} to \textbf{NM} phase for the pure case ($p=1$) gets replaced by two and three first order transitions for bimodal ($p=0$) and trimodal (fixed at $p = \frac{1}{3}$) distributions respectively.  }
\label{fig16}
\end{figure}

\item $ \mathbf{ p = \frac{1}{3}:}$ \,\,  For $0 < p < \frac{1}{3}$ regime we saw that there is always a first order transition line given by the equation  \,\,  $p \Delta +(1-p) h = \frac{1}{2}$ \,\, from \textbf{F} to the \textbf{P} phase which was bounded by two $A_5$ points ( at the junction of \textbf{P-F1-F} and \textbf{F-P-F3} phases). At exactly $p=\frac{1}{3}$, these two $A_5$ points coincide and become a $A_7$ point and the first order transition line between them vanishes. So instead of three $A_5$ points there are now two $A_7$ points and one $A_5$ point. The $A_7$ points are : one at the junction of \textbf{F-P-F1-F3} located at ($\Delta=\frac{p}{2}, \,\, h=\frac{1+p}{2}$) and  the other at the junction of the \textbf{F-NM-F1-F2} phases located at ($\Delta=\frac{1}{2}, \,\, h=\frac{1+p}{4}$). The $A_5$ point is located at the junction of \textbf{P-F1-F2} at ($\Delta=\frac{1}{2}, \,\, h=\frac{3 - p}{4}$). The phase boundary between the phase \textbf{F1} and phase \textbf{P} is given by the first order transition line with $h = \frac{1+p}{2}$ parallel to the $\Delta$ axis (see Fig. \ref{fig22}).

\item $ \mathbf{ \frac{1}{3} < p < 1:}$ \,\, For $\frac{1}{3} < p < 1$, the \textbf{F1} phase penetrates in between the phases \textbf{F-F3} and \textbf{F3-P} and the new $A_7$ point now breaks into a $A_6$ and a  $A_5$ point and a new first order transition line  $(1-p) (\Delta - h) = \frac{1}{4} (3 p^2 - 2 p -1)$ emerges, separating the phases \textbf{F1} and \textbf{F3} as shown in Fig. \ref{fig23} for $p=\frac{1}{2}$. The red square represents the $A_6$ point and it is located at ($\Delta=\frac{1-p}{4} , \,\, h=\frac{1+p}{2}$), at the junction of \textbf{F-F3-F1} phases and the new $A_5$ point is located at ($\Delta=\frac{p}{2} , \,\, h=\frac{1+ 5 p}{4}$) at the junction of \textbf{P-F3-F1} phases. The phases \textbf{F1} and \textbf{P} are again separated by the first order transition line \,\,  $(3p -1) \Delta + (1-p) h = \frac{1}{4} (1+p)^2$.

\end{itemize}

\subsubsection{$p=0 : \,\, $Bimodal distribution}

For the bimodal distribution we get the same phase diagram as obtained earlier in \cite{rfbc, santos}. The phase diagram has four phases. The phases \textbf{F1} and \textbf{F2} discussed before become a single phase which we call \textbf{F1 = F2} phase.  The other phases are the \textbf{F} phase, the \textbf{P} phase and the \textbf{NM} phase.  The first order transition lines separating these phases are similar to the ones described in the Subsection \ref{subsec3}. There are two $A_5$ points at $(\Delta=\frac{1}{2} , \, h=\frac{1}{4})$ and $(\Delta=\frac{1}{4}, \, h=\frac{1}{2})$  at the junction of \textbf{NM-F-(F1=F2)} and \textbf{F-P-(F1=F2)} phases respectively (see Fig. \ref{fig1}). 

\begin{figure}
     \centering
     \begin{subfigure}[b]{0.3\textwidth}
         \centering
         \includegraphics[width=\textwidth]{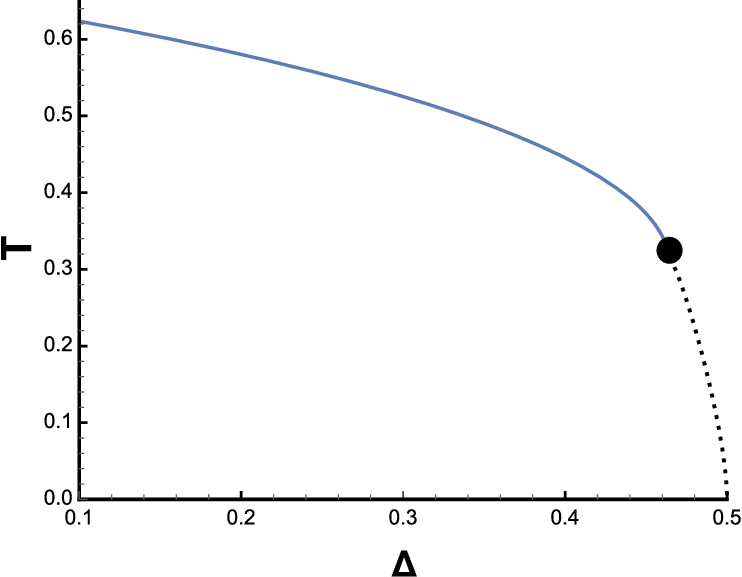}
         \caption{$ 0 \leq h < 0.257$}
         \label{fig40}
     \end{subfigure}
     \hfill
     \begin{subfigure}[b]{0.3\textwidth}
         \centering
         \includegraphics[width=\textwidth]{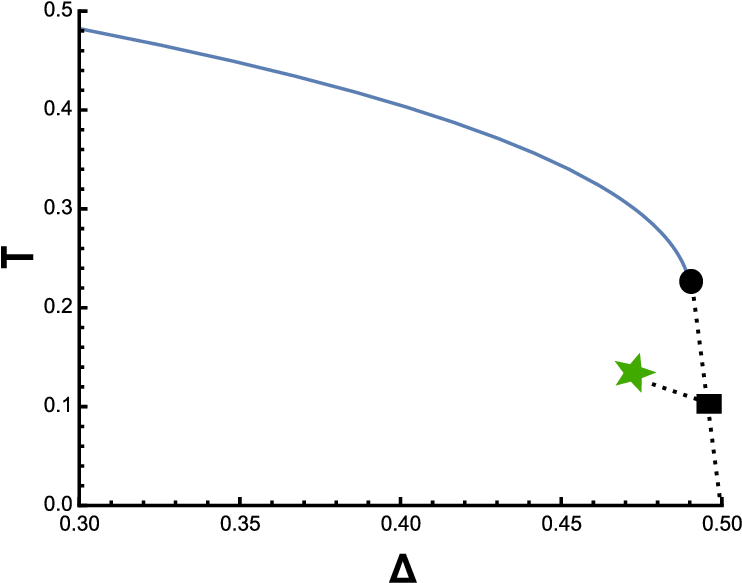}
          \caption{$ 0.257  \leq h < 0.275$}
         \label{fig41}
     \end{subfigure}
     \hfill
     \begin{subfigure}[b]{0.3\textwidth}
         \centering
         \includegraphics[width=\textwidth]{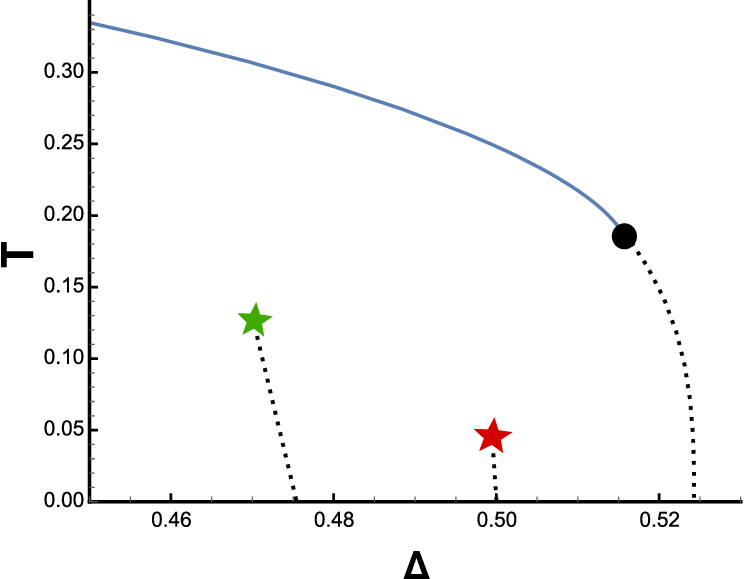}
        \caption{$ 0.275 \leq h < 0.452$}
         \label{fig42}
     \end{subfigure}
     \hfill
     \begin{subfigure}[b]{0.3\textwidth}
         \centering
         \includegraphics[width=\textwidth]{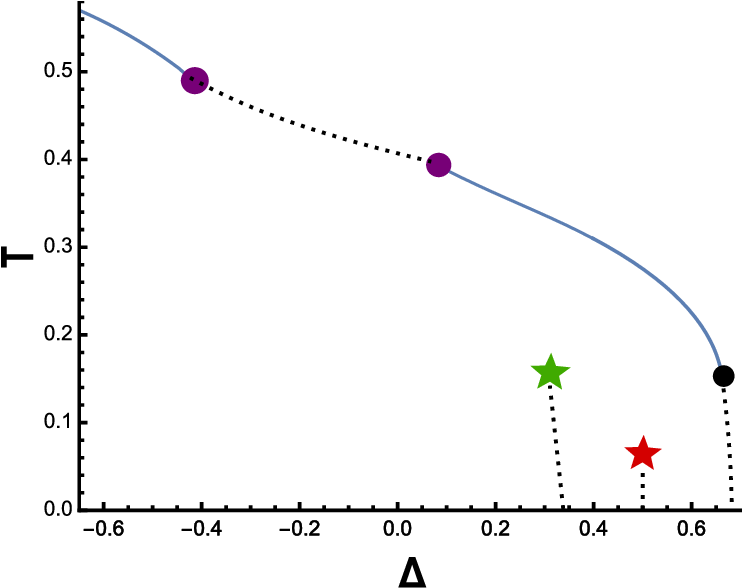}
        \caption{$ 0.452 \leq h < 0.476$}
         \label{fig43}
     \end{subfigure}
     \hfill
     \begin{subfigure}[b]{0.3\textwidth}
         \centering
         \includegraphics[width=\textwidth]{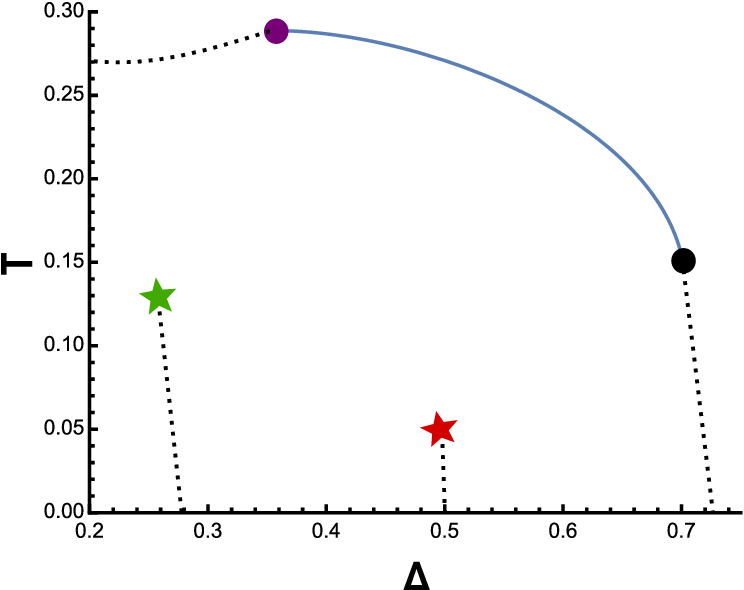}
        \caption{$ 0.476 \leq h < 0.5275$}
         \label{fig44}
     \end{subfigure}
     \hfill
     \begin{subfigure}[b]{0.3\textwidth}
         \centering
         \includegraphics[width=\textwidth]{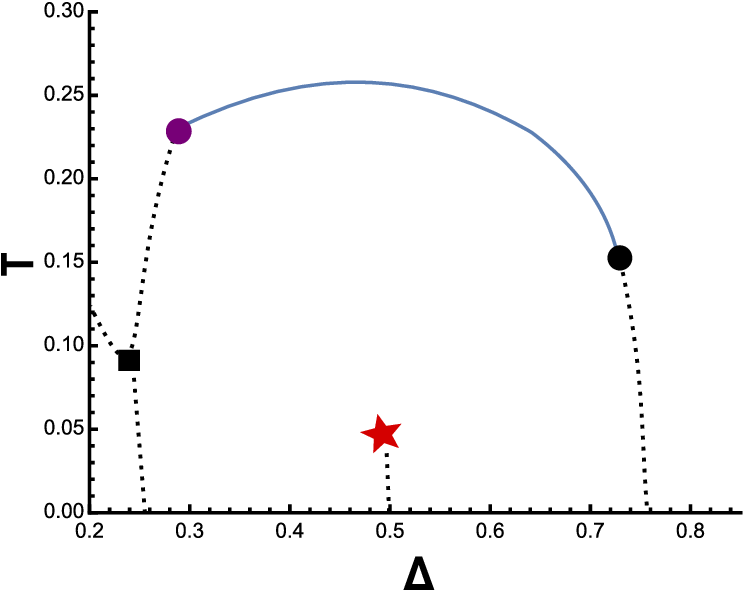}
        \caption{$ 0.5275 \leq h < 0.5281$}
         \label{fig45}
     \end{subfigure}
     \hfill
     \begin{subfigure}[b]{0.3\textwidth}
         \centering
         \includegraphics[width=\textwidth]{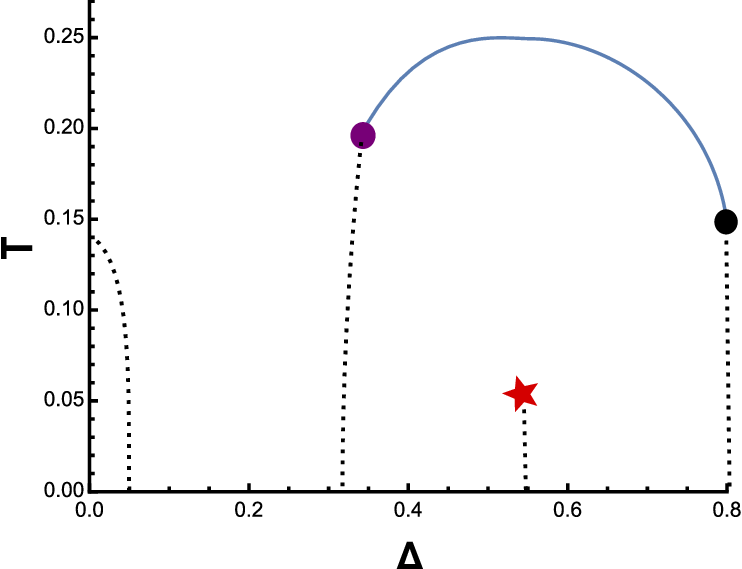}
        \caption{$ 0.5281 < h < 0.725$}
         \label{fig46}
     \end{subfigure}
     \hfill
     \begin{subfigure}[b]{0.3\textwidth}
         \centering
         \includegraphics[width=\textwidth]{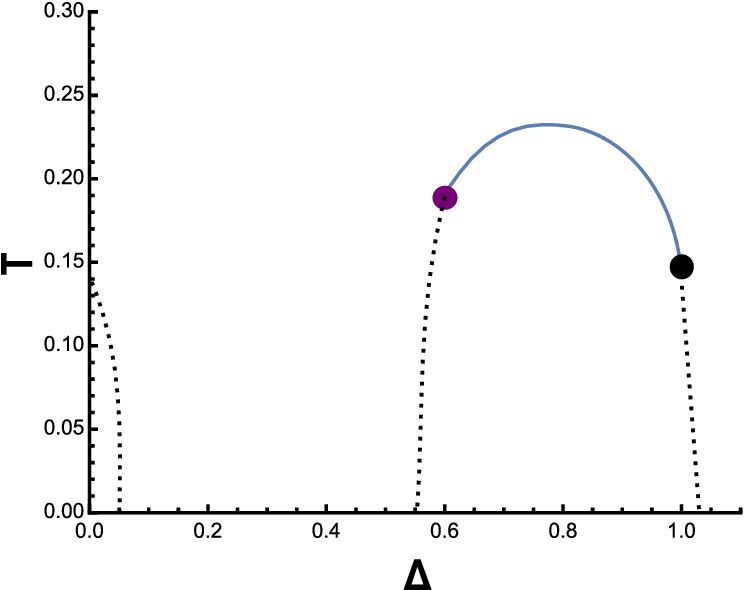}
        \caption{$  h > 0.725$}
         \label{fig47}
     \end{subfigure}
     \hfill
        \caption{ $T-\Delta$ phase diagram for different ranges of $h$ for $p=\frac{1}{10}$.  The solid line is the loci of continuous transitions and the dotted line is the loci of first order transitions,  solid stars are the BEPs,  solid circles are the TCPs,  solid squares are the $A_5$ points.  There are eight different phase diagrams  depending on the range of $h$.   }
        \label{fig4}
\end{figure}

\subsection{Gaussian distribution}\label{sec2b}

For the Gaussian distribution, the $T=0$ rate function for RFBC is :

\begin{eqnarray}\label{gaussgroundfree}
\Phi(m) & =  & \frac{ m^2}{2}  - \frac{m}{2} \Bigg (  \erf{ \Big (\frac{m + \Delta}{\sqrt{2} \sigma}} \Big )  -  \erf{ \Big (\frac{-m + \Delta}{\sqrt{2} \sigma}} \Big )   \Bigg ) +  \frac{\Delta}{2} \Bigg ( 2 -  \erf{ \Big (\frac{-m + \Delta}{\sqrt{2} \sigma}} \Big )  -  \erf{ \Big (\frac{m + \Delta}{\sqrt{2} \sigma}} \Big ) \Bigg ) \nonumber \\
		  &	- & \frac{\sigma}{ \sqrt{ 2 \pi}} \Bigg (  \exp [{- \frac{(-m + \Delta)^2}{2 \sigma^2}}]   +  \exp [{- \frac{(m + \Delta)^2}{2 \sigma^2}}]  \Bigg ) \nonumber \\
	\end{eqnarray}

Here $\erf(x)= \frac{2}{\sqrt{\pi}}  \int_{0}^{x} e^{- t^2} dt $ is the error function. The $\min\limits_m \Phi(m)$ gives the disorder averaged ground state energy. The expression of magnetization ($m$) and quadrupole moment ($q$) from Eq.  \ref{eqn1} and Eq. \ref{eqnq} after taking $\beta \rightarrow \infty$ limit are 

\begin{eqnarray}\label{magn}
m = \frac{1}{2} \Bigg (   \erf{ \Big (\frac{m + \Delta}{\sqrt{2} \sigma}} \Big )  -  \erf{ \Big (\frac{-m + \Delta}{\sqrt{2} \sigma}} \Big )   \Bigg )
\end{eqnarray}

\begin{eqnarray}\label{dens}
q = \frac{1}{2} \Bigg ( 2 -  \erf{ \Big (\frac{m + \Delta}{\sqrt{2} \sigma}} \Big )  -  \erf{ \Big (\frac{-m + \Delta}{\sqrt{2} \sigma}} \Big )   \Bigg )
\end{eqnarray}

We find that there is one ordered phase with $m \neq 0$ and one disordered phase with $m=0$. The quadrupole moment $q$ changes continuously from $q=1$ to $q=0$ as $\Delta $ goes from $0$ to $\infty$. Thus there is no transition in $q$. On expanding Eq. \ref{gaussgroundfree} around $m=0$, we get

	\begin{eqnarray}\label{landaug}
		\Phi(m) = a_2^0 m^2 + a_4^0 m^4 + a_6^0 m^6 + a_8^0 m^8+....
	\end{eqnarray}
	where, 
	
	\begin{eqnarray}
		a_2^0 &=& \frac{- \sqrt{2} e^{\frac{-\Delta^2}{2 \sigma^2}} + \sqrt{\pi} \sigma}{2 \sqrt{\pi} \sigma}\nonumber \\
		a_4^0 &=& - e^{\frac{-\Delta^2}{2 \sigma^2}}\frac{(\Delta^2 -\sigma^2)}{ 12 \sqrt{2 \pi} s^5} \nonumber \\
		a_6^0 & = & - e^{\frac{-\Delta^2}{2 \sigma^2}}\frac{(\Delta^4 -6 \Delta^2 \sigma^2 + 3 \sigma^4)}{ 360 \sqrt{2 \pi} s^9} \nonumber \\
		a_8^0  & =& - e^{\frac{-\Delta^2}{2 \sigma^2}}\frac{(\Delta^6 - 15 \Delta^4 \sigma^2 + 45 \Delta^2 \sigma^4 - 15 \sigma^6)}{ 20160 \sqrt{2 \pi} s^{13}} 	
	\end{eqnarray}
	
  \begin{figure}
\centering
\includegraphics[scale=0.7]{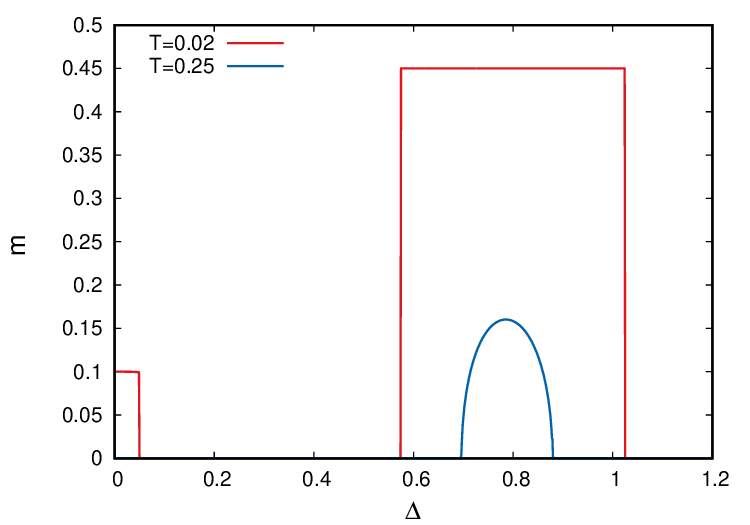}
\caption{Plot of the magnetization ($m$) as a function of $\Delta$ corresponding to the Fig. \ref{fig47} for $p=\frac{1}{10}$ at a fixed $h= 0.8$   at  two different temperatures.  At $T= 0.25$ the magnetization shows two continuous transitions, first  from \textbf{P} phase to \textbf{F2} and then to the \textbf{NM} phase.  At $T=0.02$,   for low $\Delta$   there is an ordered phase (\textbf{F3}) due to the presence of $p$ fraction of magnetic spins.   As $\Delta$ increases,  the phase undergoes a first order transition to \textbf{P} phase.  The phase \textbf{P} again undergoes a first order transition to \textbf{F2} phase which is separated from the \textbf{NM}  phase by another first order transition line.}
\label{fig150}
\end{figure}

This expansion can be used to determine the continuous transitions in the system. The line of second order transition is given by $a_2^0=0$, provided $a_4^0 >0$. This gives the line of continuous transition to be
\begin{equation}
\sigma_c = \sqrt{\frac{2}{\pi}} \exp({\frac{-\Delta_c^2}{2 \sigma_c^2}})
\end{equation}
This is valid as long as $a_4^0>0$. For $a_4^0 \leq 0$ we cannot ignore higher order terms in Eq. \ref{landaug}. We find $a_2^0=a_4^0=0$ at $\sigma_{TCP}=\Delta_{TCP}=\sqrt{\frac{2}{e \pi}} =  0.483941$. Since $a_6^0>0$ at this point, this is a tricritical point(TCP). It is shown in Fig. \ref{fig1} by a solid circle.  So the transition is second order for $ \sigma_{TCP} < \sigma \leq 1$. For $\sigma< \sigma_{TCP}$, $a_4^0<0$ and the three phases coexist. The transition becomes first order for $0 \leq \sigma < \sigma_{TCP} $ and the transition line can be found by equating the free energy and its first derivatives w.r.t $m$ on both sides. For $\sigma \rightarrow 0$,  the first order transition line cuts the $\Delta$ axis at $\Delta= \frac{1}{2}$.  The phase diagram is shown in Fig. \ref{fig3}.

\begin{figure}
     \centering
     \begin{subfigure}[b]{0.3\textwidth}
         \centering
         \includegraphics[width=\textwidth]{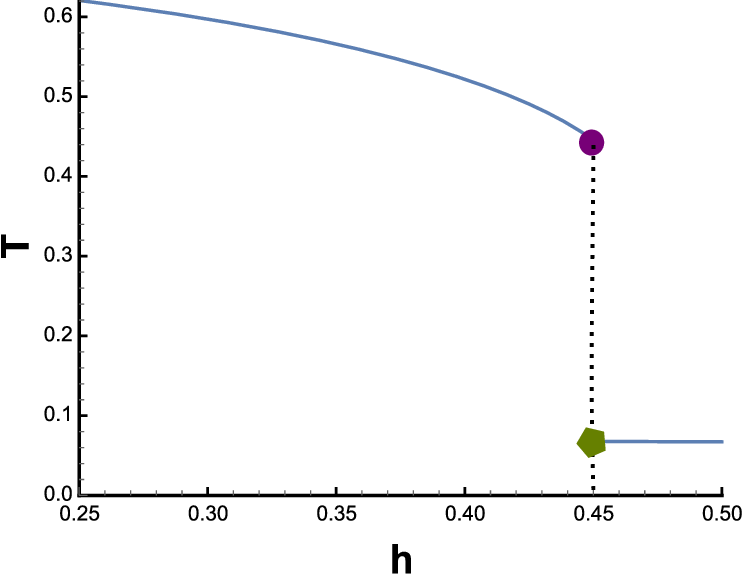}
         \caption{$- \infty < \Delta \leq 0.05$}
         \label{fig50}
     \end{subfigure}
     \hfill
     \begin{subfigure}[b]{0.3\textwidth}
         \centering
         \includegraphics[width=\textwidth]{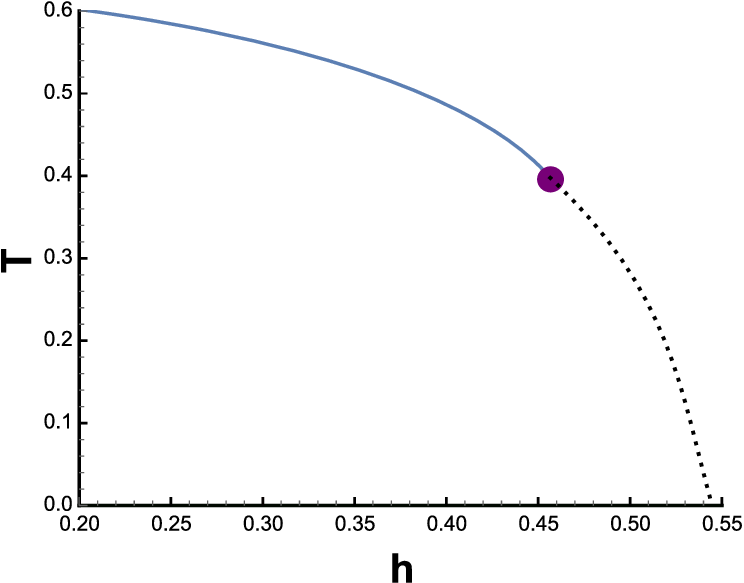}
          \caption{$ 0.05 < \Delta \leq  0.238$}
         \label{fig51}
     \end{subfigure}
     \hfill
     \begin{subfigure}[b]{0.3\textwidth}
         \centering
         \includegraphics[width=\textwidth]{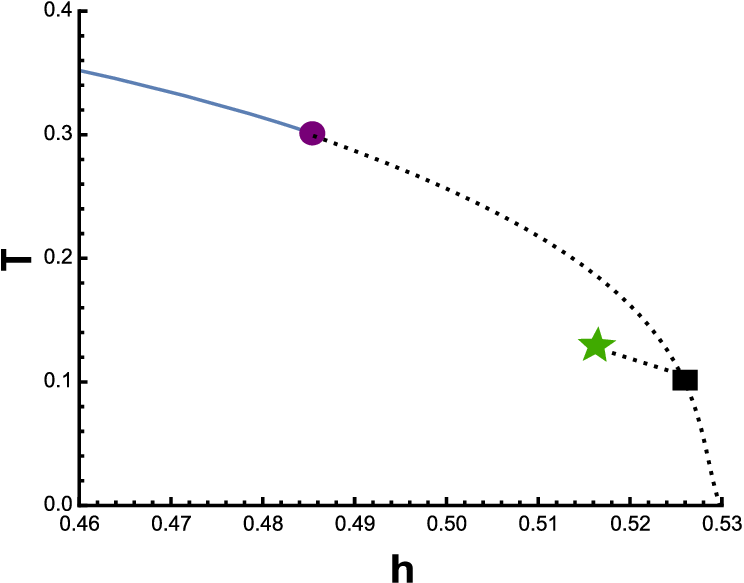}
        \caption{$ 0.238 <  \Delta < 0.247$}
         \label{fig52}
     \end{subfigure}
     \hfill
     \begin{subfigure}[b]{0.3\textwidth}
         \centering
         \includegraphics[width=\textwidth]{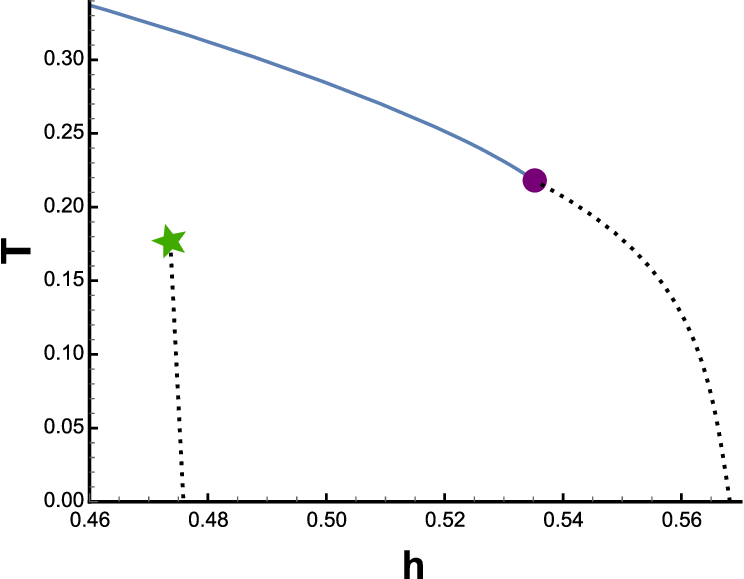}
        \caption{$ 0.247 \leq\Delta< \ln{4}/3$}
         \label{fig53}
     \end{subfigure}
     \hfill
     \begin{subfigure}[b]{0.3\textwidth}
         \centering
         \includegraphics[width=\textwidth]{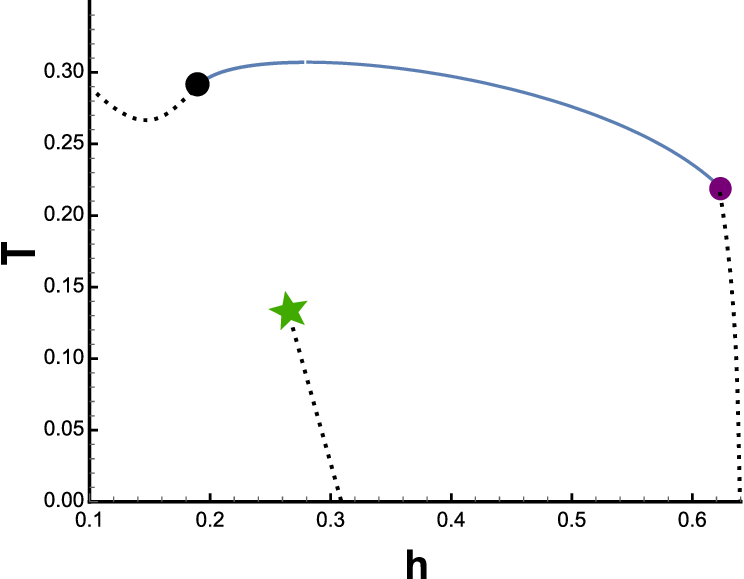}
        \caption{$ \ln{4}/3 \leq \Delta \leq 0.493$}
         \label{fig54}
     \end{subfigure}
     \hfill
     \begin{subfigure}[b]{0.3\textwidth}
         \centering
         \includegraphics[width=\textwidth]{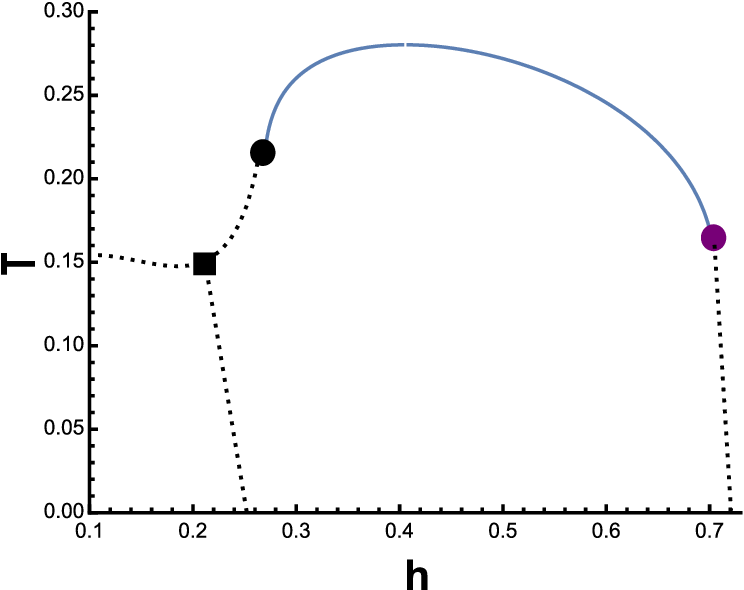}
        \caption{$ 0.493  < \Delta < \frac{1}{2}$}
         \label{fig55}
     \end{subfigure}
      \hfill
     \begin{subfigure}[b]{0.3\textwidth}
         \centering
         \includegraphics[width=\textwidth]{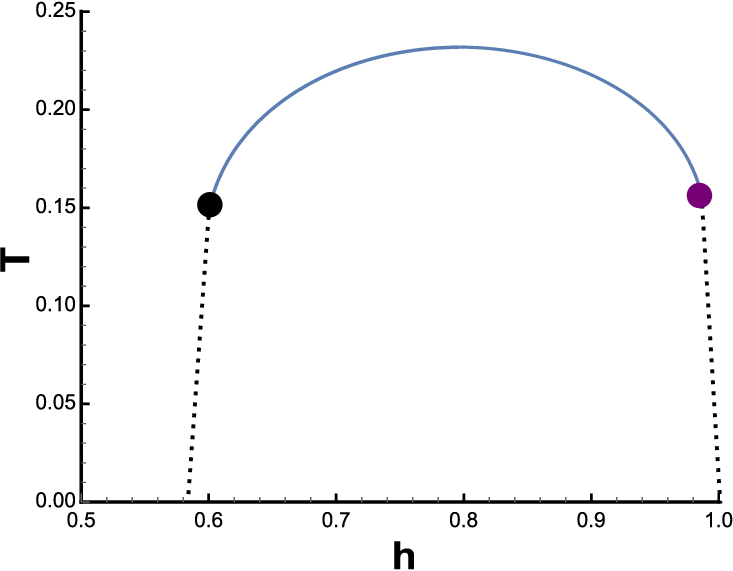}
        \caption{$ \Delta > \frac{1}{2}$}
         \label{fig56}
     \end{subfigure}
     \hfill
        \caption{ $T-h$ phase diagram for different regimes of $\Delta$ for $p=\frac{1}{10}$.  The solid line is the line of second order transitions, the dotted lines are lines of first order transition,  solid stars are the BEPs,  solid circles are the TCPs,  solid squares are the $A_5$ points and  green circles are CEPs.  There are seven different phase diagrams depending on the range of $\Delta$.   }
        \label{fig5}
\end{figure}

\section{Finite temperature phase diagram}\label{sec3}

\subsection{Trimodal distribution}\label{sec3a}

We saw that for $T=0$, there were five different phase diagrams depending on the value of $p$. One interesting and non-trivial part of the phase diagram  was the presence of multiple ordered phases separated by first order transition lines. For finite temperature, the model exhibits  phase diagrams which show re-entrance and multiple phase transitions between the ordered phases. We find that the phase diagrams can be classified into five categories just like for $T=0$. At finite temperature, multiple first order transition lines emerge separating the different ordered phases discussed in Sec. \ref{sec2}. At low temperatures, the system undergoes  two and three first order transitions as a function of both $\Delta$ and $h$ for bimodal ($p=0$) and trimodal ($0 < p < 1$) distributions respectively (see Fig. \ref{fig16} as a function of $\Delta$). Also, the system exhibits multiple TCPs. The origin of  two of them is easy to understand. One corresponds to the TCP present in the pure Blume-Capel model and the second one is the $\Delta \rightarrow -\infty$ TCP present in the RFIM with bimodal distribution. Besides these two other TCPs appear in the model. It also has BEPs and CEPs and $n$th order coexistence points denoted by $A_n$.

The magnetization in the system satisfies the fixed point equation  $\frac{\partial{f(m)}}{\partial{m}} \, = \, 0$, where $f(m)$ is the functional given by Eq. \ref{eq2}. This gives the following self-consistent equation for m

\begin{eqnarray}\label{eq3}
\frac{ m}{a}  =  \frac{1-p}{2} \Bigg( \frac{y^2 x^2-1}{y x + a y^2 x^2 + a} + \frac{x^2-y^2}{x y + a x^2+ a y^2} \Bigg )  +  p \frac{x^2-1}{x+ a x^2 +a}
\end{eqnarray}
here $a= e^{- \beta \Delta}$, $x= e^{\beta (m + H)}$ and $y= e^{\beta h}$.

Linearizing Eq. \ref{eq3} around $m=0$, we get the line of continuous transition as 
\begin{eqnarray}\label{eq4}
\frac{1}{2 \beta} = \frac{a p}{2 a + 1} + \frac{(1-p) a(2 a +  z_1)}{(1 + 2 a z_1)^2}
\end{eqnarray}
where $z_n= \cosh n\beta h$. This equation is valid only as long as the coefficients of the third order term in the expansion of Eq. \ref{eq3} in powers of $m$ is positive.

 \begin{figure}
\centering
\includegraphics[scale=0.7]{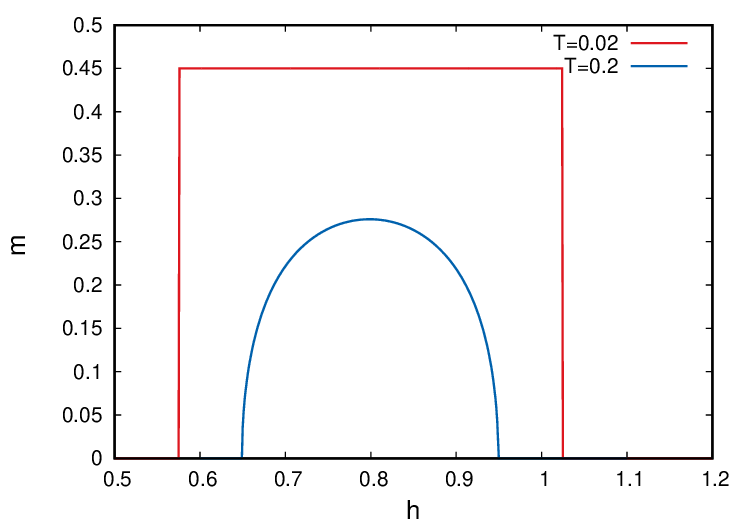}
\caption{Plot of the magnetization ($m$) as a function of $h$ corresponding to the Fig. \ref{fig56} at  $\Delta= 0.8$ and $p= \frac{1}{10}$ at two different temperatures.  At  $T= 0.2$ the magnetization undergoes two second order transitions, \textbf{NM} to \textbf{F2} phase and \textbf{F2} to \textbf{P} phase.  At $T=0.02$,  the magnetization shows two first order jumps. }
\label{fig120}
\end{figure}
 
  At TCP the line of continuous transitions (known as the $\lambda$ line) given by Eq. \ref{eq4} meets the two other lines of continuous transitions in the $T-\Delta-H$ space \cite{tcp, sumedha}.  These are the $\lambda_ \pm$ lines. At a TCP the $\lambda$, $\lambda_+$ and $\lambda_-$ lines meet in the $T-\Delta$ plane. TCP is also the end point of the $\lambda$ line given by equating the second and fourth coefficient to zero in the power series expansion of $f(m)$ for $H=0$. The BEP occurs when the $\lambda_+$ and $\lambda_-$ lines do not meet the $\lambda$ line, and instead meet at a point in the ordered region in the $T-\Delta$ plane. In order to locate the BEP we use the general condition of criticality by equating the first three derivatives of the free energy to  zero ($f'(m) = f''(m) = f'''(m) =0$) along with the condition, $f''''(m) > 0$. We get the following two equations by equating $f''(m)=0$ and  $f'''(m) =0$ respectively

\begin{eqnarray}\label{eq7}
 \frac{1-p}{2} \Bigg ( \frac{  4 a y x + y^2 x^2+1}{(y x + a y^2 x^2 + a)^2} &+& \frac{  4 a x y + x^2+y^2}{(x y + a x^2+ a y^2)^2} \Bigg ) y   + p \,\,  \frac{4 a x +x^2+1}{(x+ a x^2 +a)^2} \,=  \frac{1 }{ \beta a x} 
\end{eqnarray}

\begin{eqnarray}\label{eq8}
\frac{1-p}{2} \Bigg ( \frac{ (x y   - 8 a^2 y x -  a y^2 x^2 -a )   (x^2 y^2 -1)}{(y x + a y^2 x^2 + a)^3}  +   \frac{ (x y - 8 a^2 y x - a  x^2 -a y^2 ) (x^2  -y^2)}{(y x + a  x^2 + a y^2)^3} \Bigg ) y   
  + p \,\, \frac{ (x  - 8 a^2  x - a  x^2 -a ) (x^2  -1)}{( x + a  x^2 + a)^3} =0 \nonumber \\ 
\end{eqnarray}

 Numerically solving Eq.\ref{eq3}, Eq.\ref{eq7} and Eq.\ref{eq8}  simultaneously for $T$, $\Delta$ and $h$ by taking $H=0$ we find the coordinates of the point of intersection of the $\lambda_\pm$ lines. If $m=0$ at the point of intersection, then it is a TCP, else it is a BEP.

\begin{figure}
     \centering
     \begin{subfigure}[b]{0.3\textwidth}
         \centering
         \includegraphics[width=\textwidth]{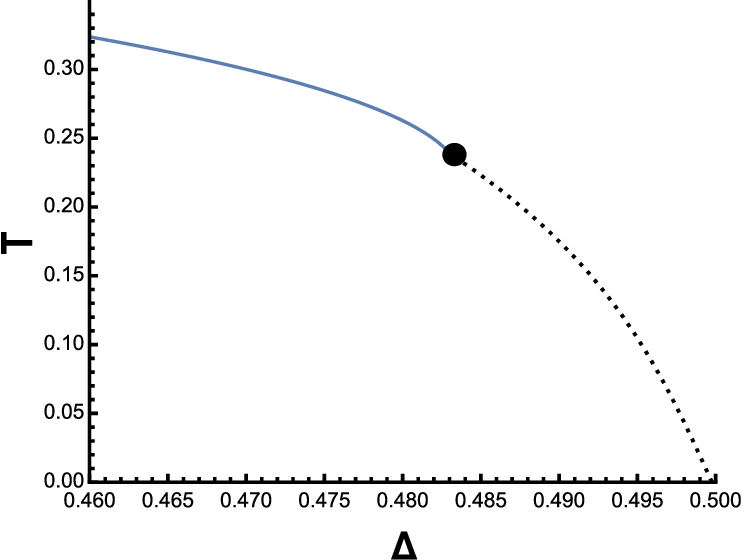}
         \caption{$ 0 \leq h < 0.325$}
         \label{fig90}
     \end{subfigure}
     \hfill
     \begin{subfigure}[b]{0.3\textwidth}
         \centering
         \includegraphics[width=\textwidth]{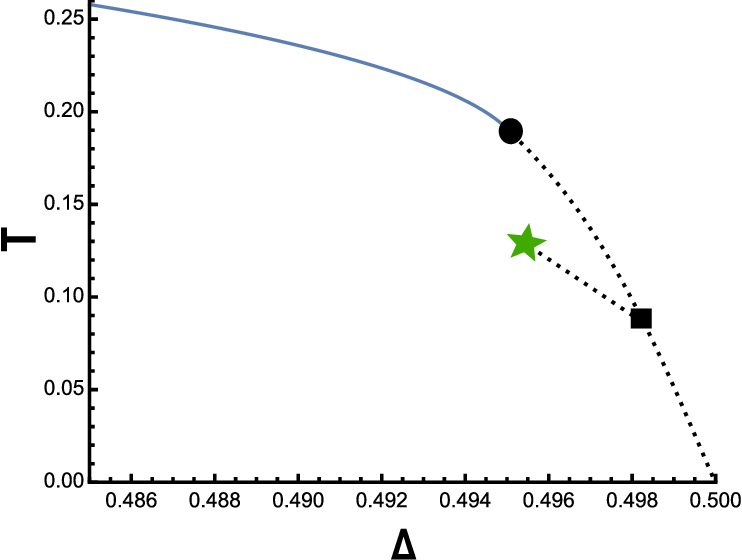}
          \caption{$ 0.325 \leq h < 0.33$}
         \label{fig91}
     \end{subfigure}
     \hfill
     \begin{subfigure}[b]{0.3\textwidth}
         \centering
         \includegraphics[width=\textwidth]{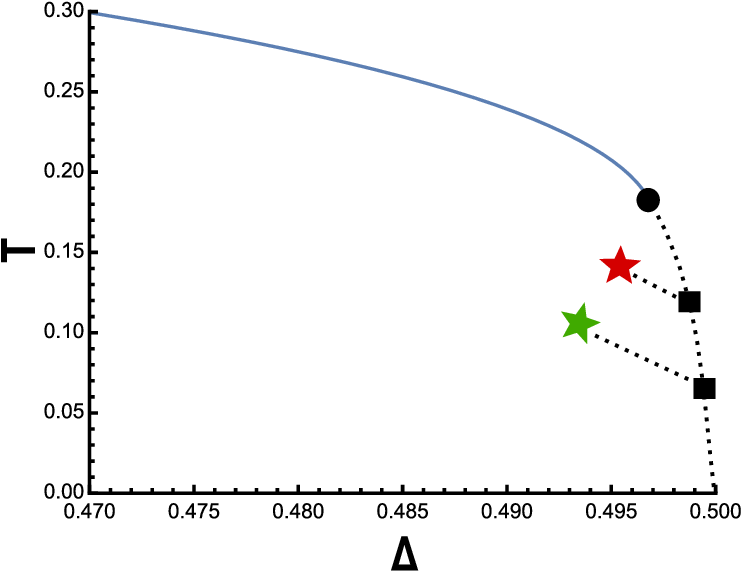}
        \caption{$ 0.33 \leq h < 0.333$}
         \label{fig92}
     \end{subfigure}
     \hfill
     \begin{subfigure}[b]{0.3\textwidth}
         \centering
         \includegraphics[width=\textwidth]{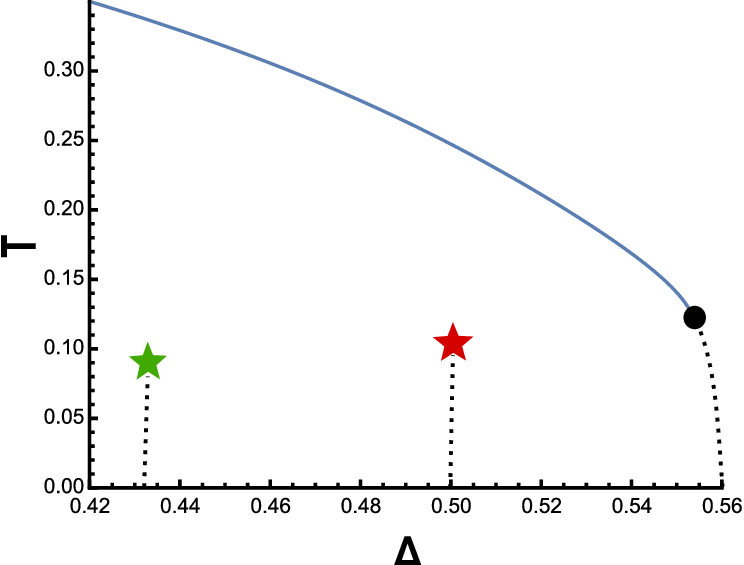}
        \caption{$ 0.333 \leq h < 0.6079$}
         \label{fig93}
     \end{subfigure}
     \hfill
     \begin{subfigure}[b]{0.3\textwidth}
         \centering
         \includegraphics[width=\textwidth]{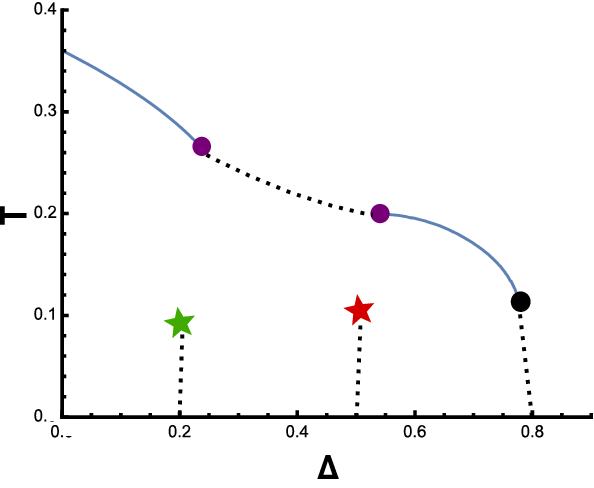}
        \caption{$ 0.6079 \leq h \leq 0.63$}
         \label{fig95}
     \end{subfigure}
     \hfill
      \begin{subfigure}[b]{0.3\textwidth}
         \centering
         \includegraphics[width=\textwidth]{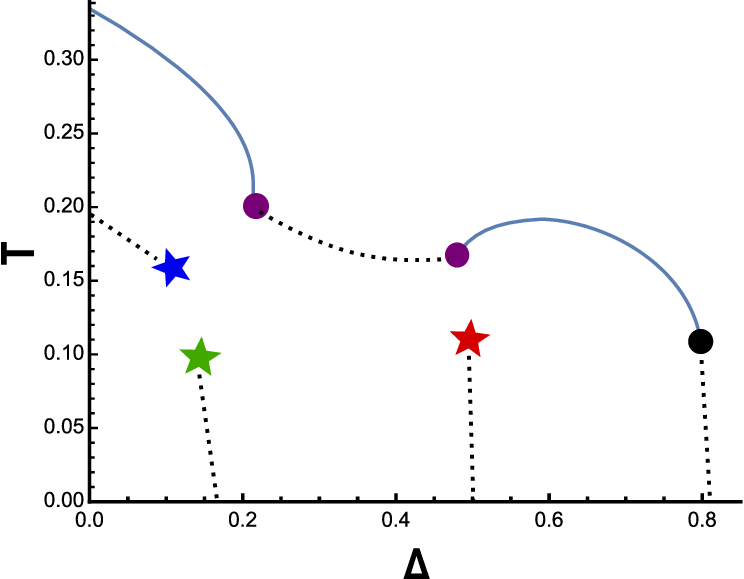}
        \caption{$ 0.63 < h < 0.658$}
         \label{fig99}
     \end{subfigure}
     \hfill
     \begin{subfigure}[b]{0.3\textwidth}
         \centering
         \includegraphics[width=\textwidth]{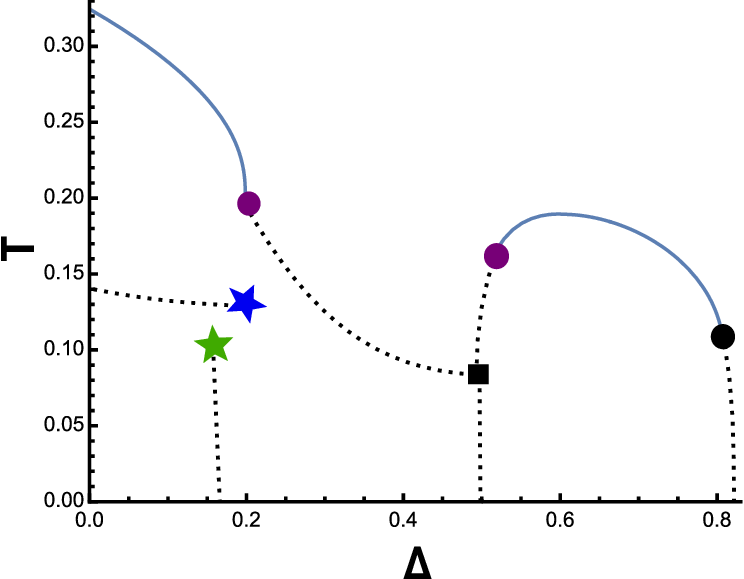}
        \caption{$ 0.658 \leq h < 0.66$}
         \label{fig96}
     \end{subfigure}
     \hfill
     \begin{subfigure}[b]{0.3\textwidth}
         \centering
         \includegraphics[width=\textwidth]{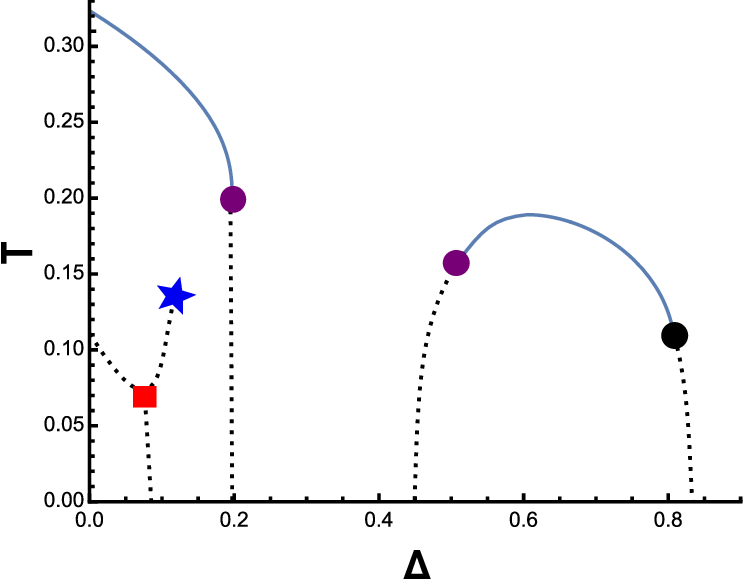}
        \caption{$ 0.66 \leq h < 0.666$}
         \label{fig97}
     \end{subfigure}
     \hfill
     \begin{subfigure}[b]{0.3\textwidth}
         \centering
         \includegraphics[width=\textwidth]{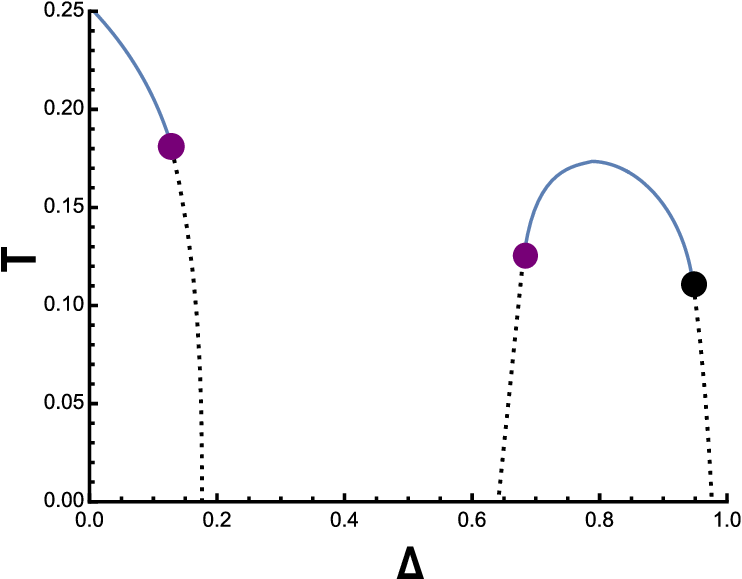}
        \caption{$  h > 0.666$}
         \label{fig98}
     \end{subfigure}
     \hfill
        \caption{ $T-\Delta$ phase diagram for different regions of $h$ for $p=\frac{1}{3}$.  The solid line are the lines of second order transitions, the dotted lines are the lines of first order transitions,  solid stars are the BEPs,  solid circles are the TCPs and  solid squares are  the $A_5$ points.  There are nine different phase diagrams  depending on the range of $h$.  }
        \label{fig9}
\end{figure}

We have studied the entire range of $p$ and we find many different phases, depending on the values of $T$, $\Delta$ and $h$. We give the details of all possible phase diagrams in this section.

 \subsubsection{Phase diagram of pure Blume-Capel model : $p=1$}

 The phase diagram of the pure Blume-Capel model in the $T-\Delta$ plane is well known \cite{beg}. It is similar to Fig. \ref{fig3}, with $x-$axis being the temperature $T$. There is a line of second order transition which meets the line of first order transition at a TCP  ($T_{BC}= \frac{1}{3}, \,\,\, \Delta_{BC}= \frac{\ln 4}{3}$). The first order transition line continues till $T=0$, with the first order transition at $\Delta = \frac{1}{2}$ for $T=0$.

\subsubsection{Phase diagrams for $p=\frac{1}{10}$}

\begin{figure}
     \centering
     \begin{subfigure}[b]{0.3\textwidth}
         \centering
         \includegraphics[width=\textwidth]{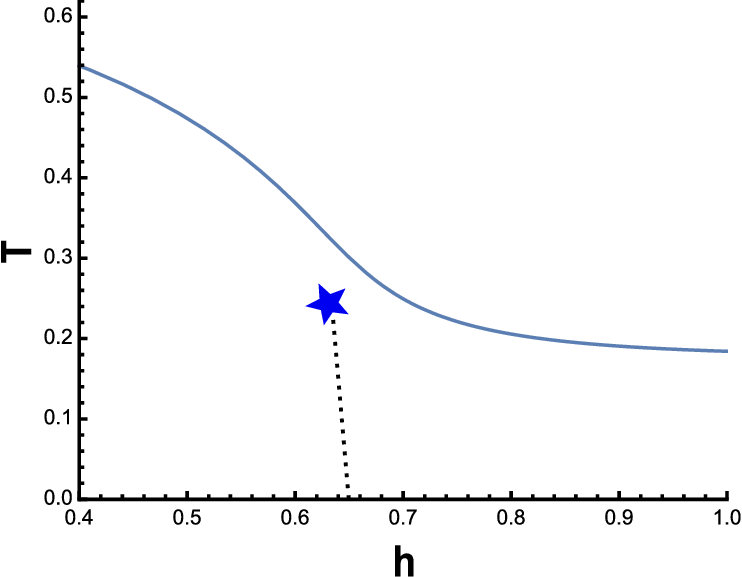}
         \caption{$- \infty < \Delta \leq 0.154$}
         \label{fig100}
     \end{subfigure}
     \hfill
     \begin{subfigure}[b]{0.3\textwidth}
         \centering
         \includegraphics[width=\textwidth]{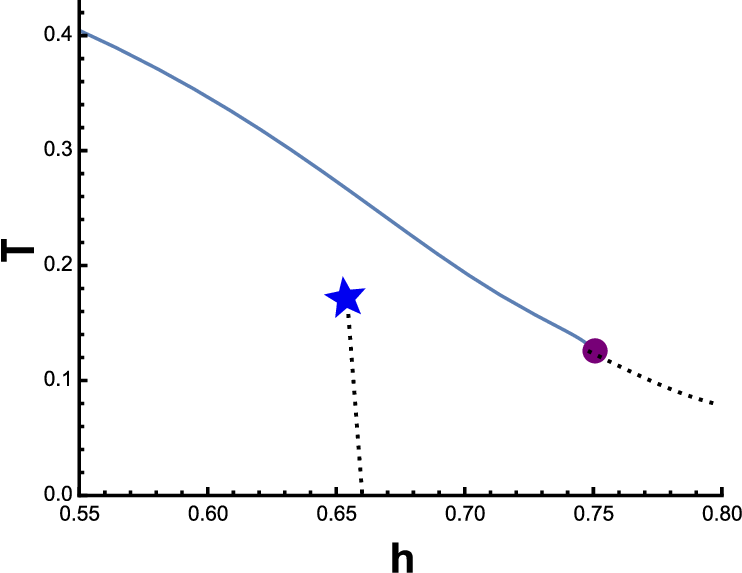}
          \caption{$0.154 < \Delta < 0.163$}
         \label{fig101}
     \end{subfigure}
     \hfill
     \begin{subfigure}[b]{0.3\textwidth}
         \centering
         \includegraphics[width=\textwidth]{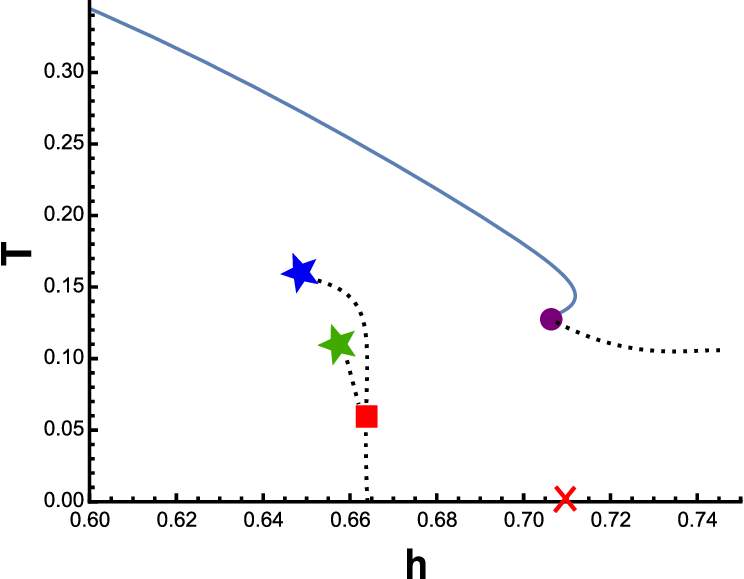}
        \caption{$0.163 \leq \Delta < 0.167$}
         \label{fig102}
     \end{subfigure}
     \hfill
     \begin{subfigure}[b]{0.3\textwidth}
         \centering
         \includegraphics[width=\textwidth]{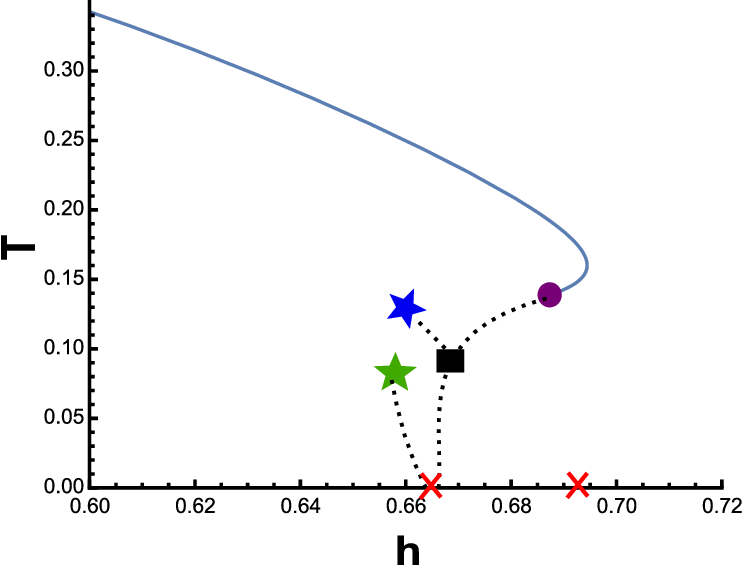}
        \caption{$0.167 \leq \Delta < 0.18$}
         \label{fig103}
     \end{subfigure}
     \hfill
     \begin{subfigure}[b]{0.3\textwidth}
         \centering
         \includegraphics[width=\textwidth]{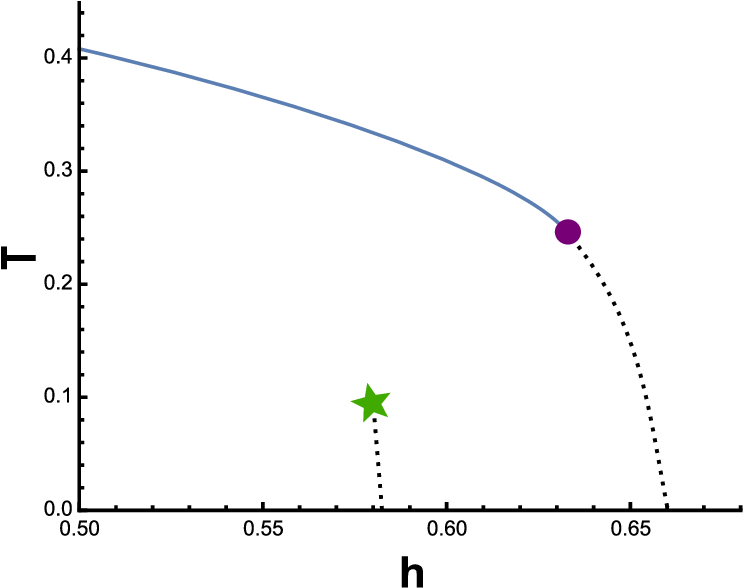}
        \caption{$0.18 \leq \Delta < \ln{4}/3$}
         \label{fig104}
     \end{subfigure}
     \hfill
     \begin{subfigure}[b]{0.3\textwidth}
         \centering
         \includegraphics[width=\textwidth]{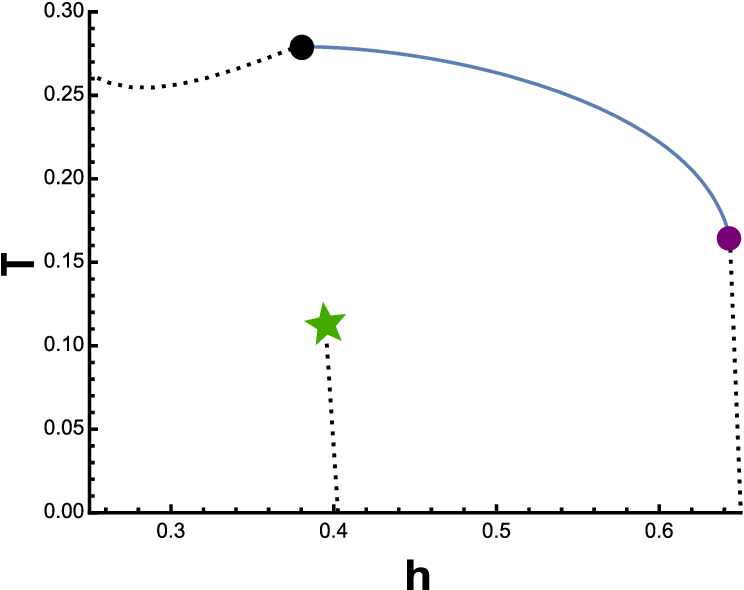}
        \caption{$\ln{4}/3 \leq \Delta < 0.498$}
         \label{fig105}
     \end{subfigure}
     \hfill
     \begin{subfigure}[b]{0.3\textwidth}
         \centering
         \includegraphics[width=\textwidth]{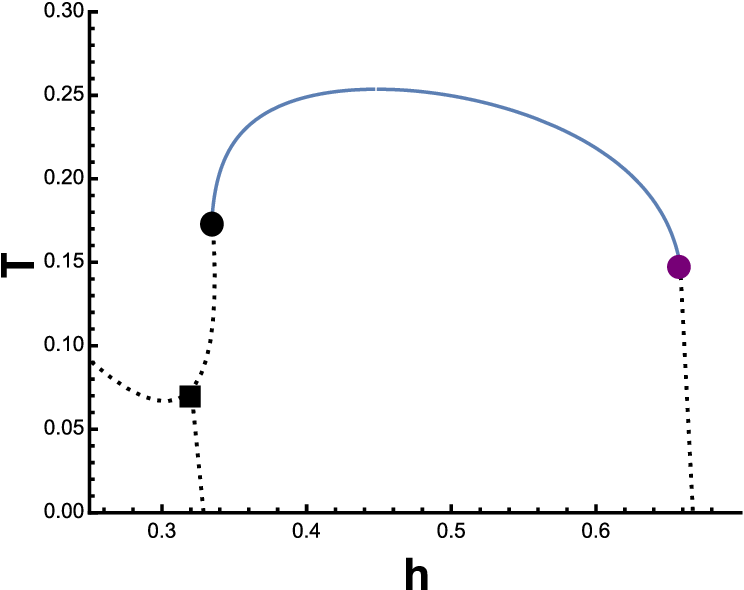}
        \caption{$0. 498 \leq \Delta < \frac{1}{2}$}
         \label{fig106}
     \end{subfigure}
     \hfill
     \begin{subfigure}[b]{0.3\textwidth}
         \centering
         \includegraphics[width=\textwidth]{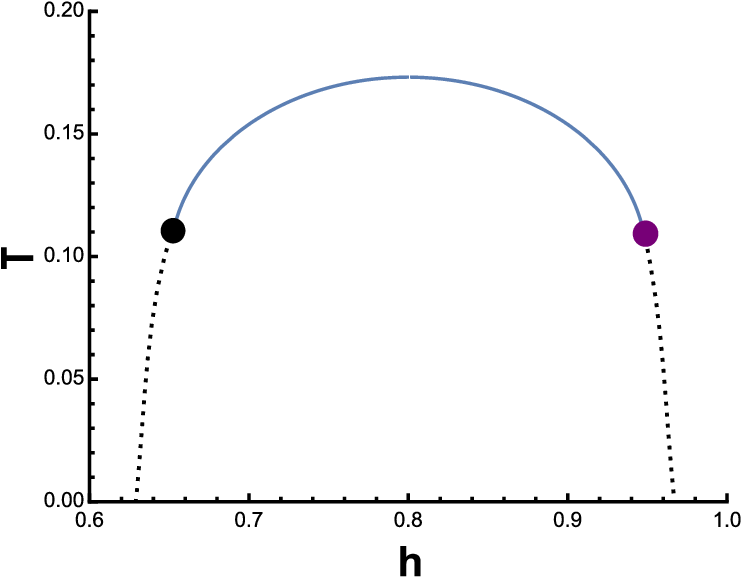}
        \caption{$\Delta > \frac{1}{2}$}
         \label{fig107}
     \end{subfigure}
     \hfill
        \caption{ $T-h$ phase diagram for different regimes of $\Delta$ for $p=\frac{1}{3}$.  The solid  lines are the lines of  second order transitions, the dotted lines are first order transitions, the  solid stars are the BEPs, the solid circles are the TCPs,  solid squares are the $A_5$ points and  red squares are the $A_6$ points.  There are eight different phase diagrams  depending on the range of $\Delta$.   }
        \label{fig10}
\end{figure}

  Fig. \ref{fig4} shows  different phase diagrams in the $T-\Delta$ plane for different ranges of $h$ for $p=\frac{1}{10}$.  There are eight  different phase diagrams depending on the value of $h$. For $0 \leq h < 0.257$, the phase diagram is similar to the pure model (see Fig.\ref{fig40}). As $h$ increases, new multicritical points arise. For $0.257 \leq h < 0.275$, another first order transition line emerges (shown by dotted lines) separating \textbf{F-F1} phases, which ends at a BEP. The two first order transition lines meet at a $A_5$ point (Fig. \ref{fig41}). For $0.275 \leq h < 0.452$, the phase diagram consists of three first order transition lines separating \textbf{F-F1}, \textbf{F1-F2} and \textbf{F2-NM} phases respectively as $\Delta$ increases (see Fig. \ref{fig42}). For $0.452 \leq h < 0.476$ (shown in Fig. \ref{fig43}), the $\lambda$ line (shown by a continuous line) separates into two parts, which are connected by a first order transition line. This gives rise to three TCPs in the system. As we  increase $h$ further, one of the TCP vanishes  and the phase diagram consists of two TCPs and two BEPs (see Fig. \ref{fig44}). For $0.5275 \leq h < 0.5281$ one of the BEP gets replaced by a $A_5$ point (Fig. \ref{fig45}).

  At $\Delta=0.5281$ the $A_5$ point moves to $T=0$ and the phase diagram divides into two parts. The ordered phase \textbf{F3} exists for small $\Delta$. And for large $\Delta$ the phases \textbf{F1} and \textbf{F2} are separated by a first order transition line which ends at a BEP. These phases are bounded by the two disordered phases : for higher $\Delta$ the phase is \textbf{NM} and the intermediate disordered phase between the two parts is \textbf{P} (Fig. \ref{fig46}). For $h > 0.725$, the BEP vanishes and the phase diagram contains two TCPs (see Fig. \ref{fig47}). The plot of the magnetization for Fig. \ref{fig47} is shown in  Fig. \ref{fig150}.

 The $T-h$ phase diagrams for different values of $\Delta$ are shown in Fig. \ref{fig5}. There are seven different phase diagrams depending on the value of $\Delta$.  For $-\infty < \Delta < 0.05$, the phase diagram consists two lines of continuous transition, a TCP and a CEP. The \textbf{F3} phase occurs at low $T$ bounded by a line of second order transitions. This second order transition line meets the first order transition  line at a CEP (shown in Fig. \ref{fig50}). CEP is a  point where a second order transition line abruptly terminates onto a first order transition line. As $\Delta$ increases, the CEP vanishes (Fig. \ref{fig51}). On increasing $\Delta$ further, a first order transition line arises separating \textbf{F-F1} phases and ends at a BEP (Fig. \ref{fig52}, \ref{fig53}). At exactly $\Delta = \Delta_{BC}= \ln 4 / 3$, another TCP emerges at $\beta = 3$ and $h=0$, corresponding to the TCP of the pure model. For $  \Delta_{BC} \leq \Delta \leq 0.493$, there are two TCPs and one BEP (see Fig. \ref{fig54}). As $\Delta$ increases further, the BEP turns into a $A_5$ point (Fig. \ref{fig55}).

  Fig.  \ref{fig56} is the phase diagram for  $\Delta>0.5$. There is only one ordered phase \textbf{F2} which exists for high values of $h$. This phase is separated from the two disordered phases by two first order transition lines with  \textbf{P} phase for higher $h$ and  \textbf{NM} phase for lower $h$. The behaviour of the magnetization for some fixed values of $T$ along the $h$ axis is shown in Fig. \ref{fig120}.

 For all $p < \frac{1}{3}$ we find similar phase diagrams. Although, depending on $p$, the exact location of the transitions for different phase diagram changes.

\subsubsection{Phase diagrams for $p=\frac{1}{3}$}

\begin{figure}
     \centering
     \begin{subfigure}[b]{0.32\textwidth}
         \centering
         \includegraphics[width=\textwidth]{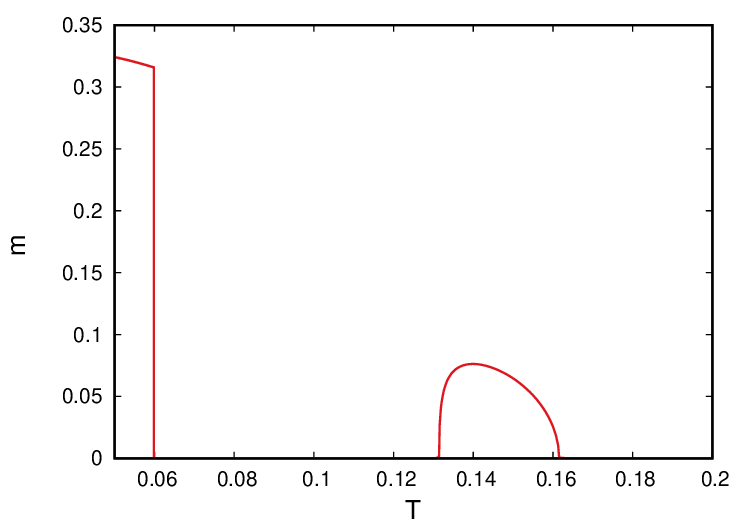}
         \caption{}
         \label{fig990}
     \end{subfigure}
     \hfill
     \begin{subfigure}[b]{0.32\textwidth}
         \centering
         \includegraphics[width=\textwidth]{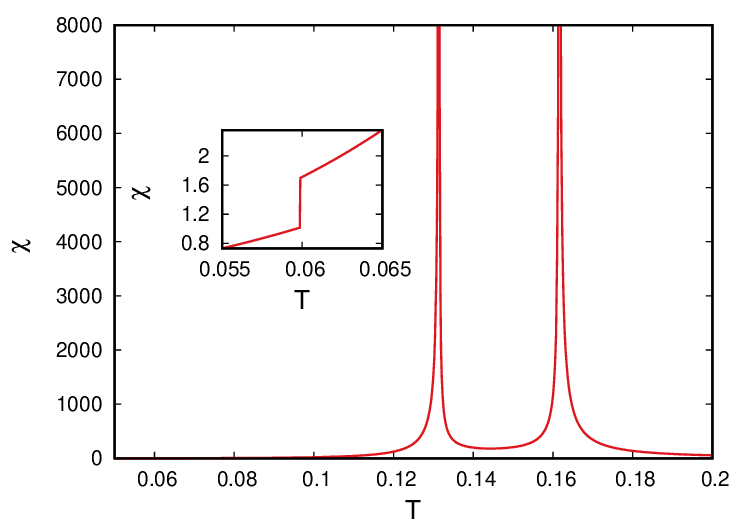}
          \caption{}
         \label{fig991}
     \end{subfigure}
     \hfill
     \begin{subfigure}[b]{0.32\textwidth}
         \centering
         \includegraphics[width=\textwidth]{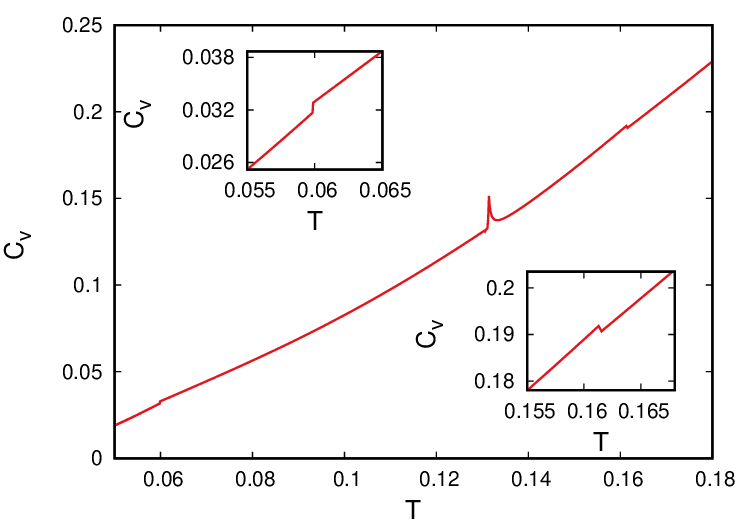}
        \caption{}
         \label{fig992}
     \end{subfigure}
     \hfill
        \caption{ Thermodynamic quantities near the re-entrance regime of Fig. \ref{fig102} for $p=\frac{1}{3}$  at $\Delta=0.165$. Fig. \textbf{(a)} shows the magnetization ($m$) as a function of $T$ for $h=0.708$.  The magnetization at the second order transition at $T_c= 0.131407$ fits with the scaling function $ 2.008 \mid T-T_c \mid^{0.5}$ and the second transition at $T_c= 0.1615509$ fits with the scaling function $ 0.628833 \mid T-T_c \mid^{0.5}$. Fig. \textbf{(b)} is the susceptibility ($\chi$) plot. $\chi$ shows two divergences at the two continuous transition points and the inset shows the discontinuity in $\chi$ at the low $T$ due to the first order transition. Fig. \textbf{(c)} is the plot of the specific heat ($C_v$). There are three jumps in the $C_v$ plot at the three transition points. }
        \label{fig999}
\end{figure}

  Fig. \ref{fig9} shows the different phase diagrams in the $T-\Delta$ plane for different ranges of $h$ for $p=\frac{1}{3}$. There are now nine different phase diagrams. Four of the phase diagrams (Fig. \ref{fig90}, Fig.\ref{fig91}, Fig. \ref{fig93} and Fig. \ref{fig95}) are similar to the phase diagrams for $p=\frac{1}{10}$ (Fig. \ref{fig40} - Fig. \ref{fig43}). In the intermediate values of $h$ between Fig. \ref{fig91} and Fig. \ref{fig93}, the phase diagram has three first order lines, two of them are inside the ordered region separating the phases \textbf{F-F1} and \textbf{F1-F2}. These two lines start at different $A_5$ points and end at two different BEPs (see Fig. \ref{fig92}). For $0.63 < h < 0.658$, the phase diagram consists of three BEPs and two TCPs, see Fig. \ref{fig99}. As $h$ increases, one BEP turns into a $A_5$ point (Fig. \ref{fig96}) and as $h$ increases further another BEP turns into a $A_6$ point (Fig. \ref{fig97}). Finally, for all $h > \frac{2}{3}$, there is always an ordered phase \textbf{F2} for large $\Delta$ separated from the disordered phases by two first order transition lines, and another ordered phase \textbf{F3} for small $\Delta$. Thus there are three TCPs in this range of $h$ (see Fig. \ref{fig98}).

\begin{figure}
     \centering
     \begin{subfigure}[b]{0.48\textwidth}
         \centering
         \includegraphics[width=\textwidth]{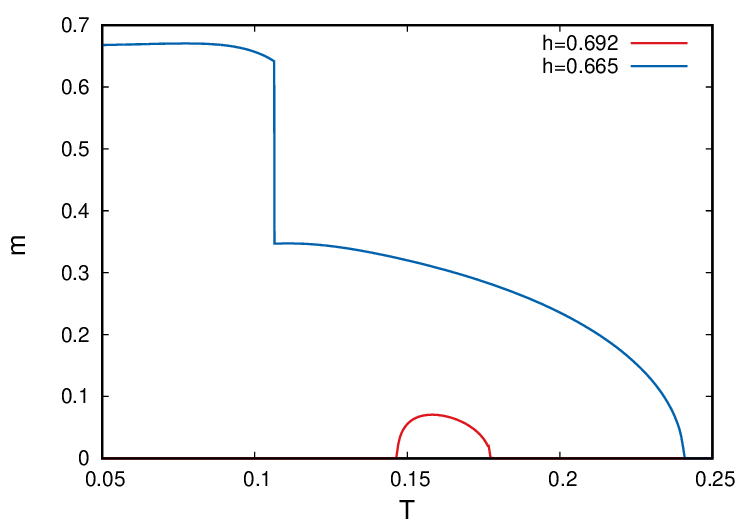}
         \caption{}
         \label{fig880}
     \end{subfigure}
     \hfill
     \begin{subfigure}[b]{0.48\textwidth}
         \centering
         \includegraphics[width=\textwidth]{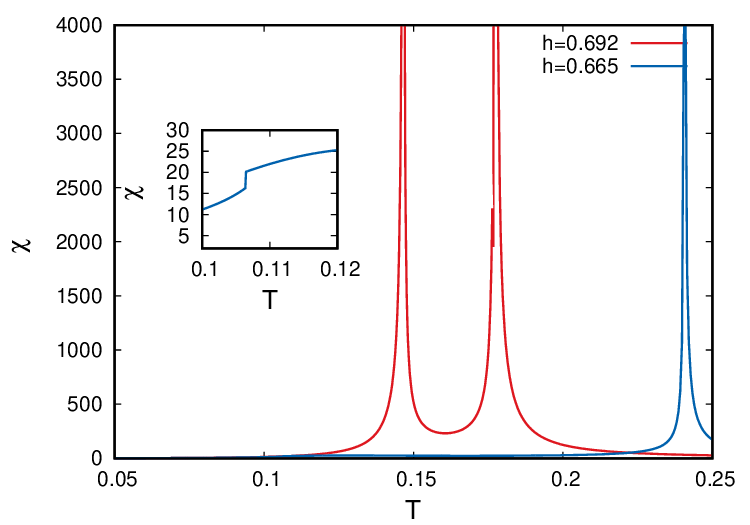}
          \caption{}
         \label{fig881}
     \end{subfigure}
     \hfill
     \begin{subfigure}[b]{0.48\textwidth}
         \centering
         \includegraphics[width=\textwidth]{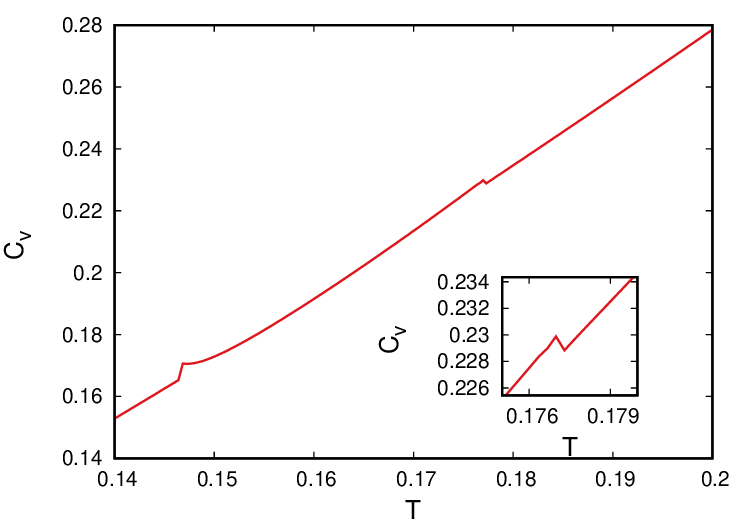}
        \caption{}
         \label{fig882}
     \end{subfigure}
     \hfill
     \begin{subfigure}[b]{0.48\textwidth}
         \centering
         \includegraphics[width=\textwidth]{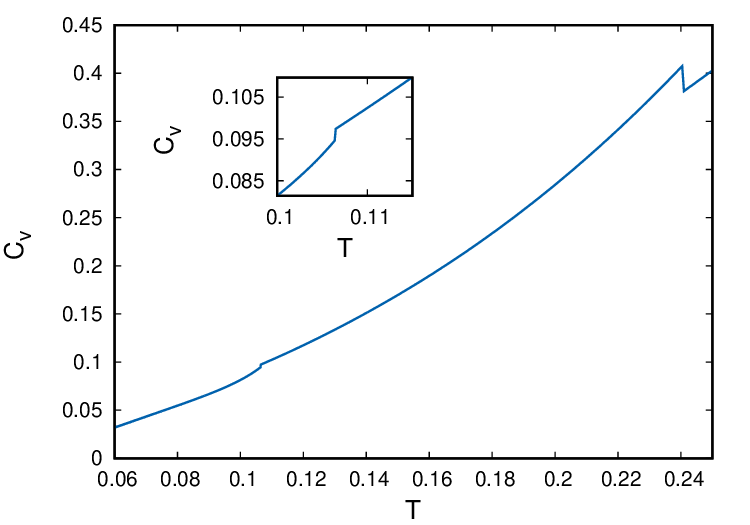}
        \caption{}
         \label{fig883}
     \end{subfigure}
     \hfill
        \caption{ Thermodynamic quantities near the re-entrance regime of Fig. \ref{fig103} for $p=\frac{1}{3}$  at $\Delta=0.17$. The blue solid line is for $h=0.692$ and red solid line is for $h=0.665$. Fig. \textbf{(a)} shows the magnetization ($m$). At $h=0.692$, the phase diagram shows re-entrance in $m$ and for $h=0.665$  $m$ has two transitions : a first order transition at lower temperature and a continuous transition at higher temperature. Fig. \textbf{(b)} the susceptibility ($\chi$) is plotted for both the values of $h$. It confirms the nature of transition in $m$. Fig. \textbf{(c)} and Fig. \textbf{(d)}  are the plot of the specific heat ($C_v$) for $h=0.692$ and $h=0.665$ respectively. $C_v$ is discontinuous at all transition points.}
        \label{fig88}
\end{figure}

    Similarly, the projection of the phase diagrams in the $T-h$ plane can be divided into eight categories depending on the ranges of $\Delta$, shown in Fig. \ref{fig10}. Four of the phase diagrams (Fig. \ref{fig104}, Fig. \ref{fig105}, Fig. \ref{fig106} and Fig. \ref{fig107}) are similar to the $T-h$ phase diagrams of the $p =\frac{1}{10}$ case (Fig. \ref{fig53}, Fig. \ref{fig54}, Fig. \ref{fig55} and Fig. \ref{fig56}). For $- \infty < \Delta \leq 0.154$, there is one first order transition  line separating \textbf{F-F3 } phases ending at a BEP. The \textbf{F3} phase exists for all values of $h$ (Fig.\ref{fig100}). For $0.154 < \Delta < 0.163$, a TCP arises (Fig. \ref{fig101}). For $0.163 \leq \Delta <  0.167$ another first order transition line appears separating \textbf{F1-F3} phases, and the two first order transition  lines meet at a $A_6$ point, see Fig. \ref{fig102}. As $\Delta$ increases further, two of the first order transition lines meet at a $A_5$ point and the phase diagram now has no ordered phases at low $T$ for high values of $h$ (Fig. \ref{fig103}).

We find re-entrance in the phase diagrams shown in  Fig. \ref{fig102} and Fig. \ref{fig103}. We have studied the magnetization, susceptibility, free energy and specific heat in the re-entrance region. The details are given in  the next subsection.

\paragraph{Re-entrance for $p=\frac{1}{3}$}\label{appdx2}

\begin{figure}
     \centering
     \begin{subfigure}[b]{0.3\textwidth}
         \centering
         \includegraphics[width=\textwidth]{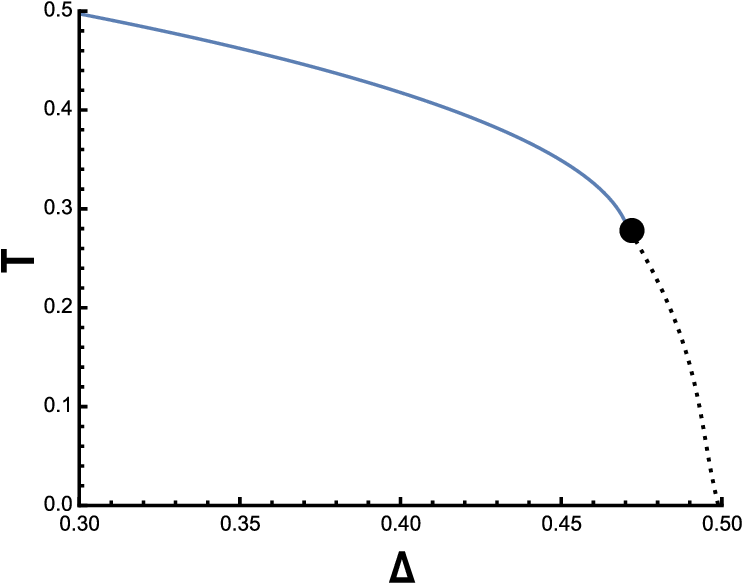}
         \caption{$ 0 \leq h < 0.374$}
         \label{fig60}
     \end{subfigure}
     \hfill
     \begin{subfigure}[b]{0.3\textwidth}
         \centering
         \includegraphics[width=\textwidth]{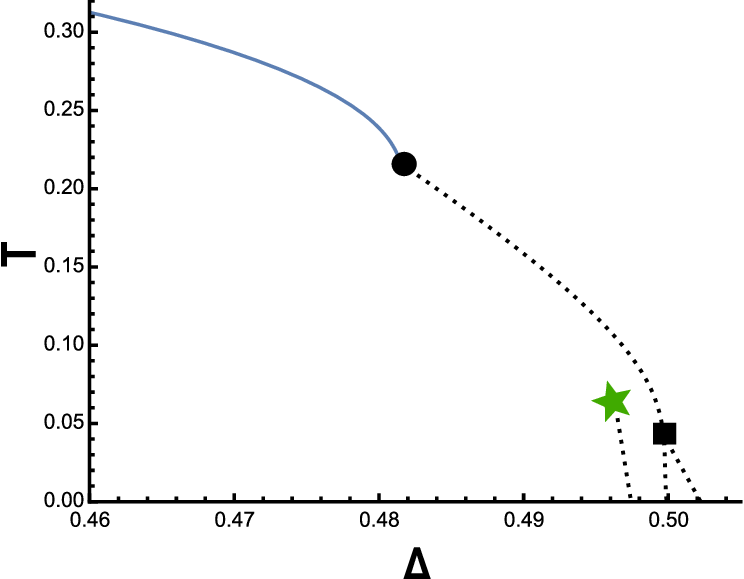}
          \caption{$ 0.374  \leq h < 0.378$}
         \label{fig61}
     \end{subfigure}
     \hfill
     \begin{subfigure}[b]{0.3\textwidth}
         \centering
         \includegraphics[width=\textwidth]{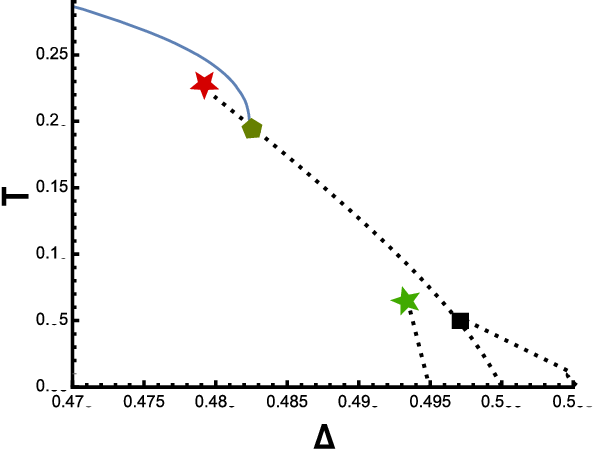}
        \caption{$ 0.378 \leq h < 0.4$}
         \label{fig62}
     \end{subfigure}
     \hfill
     \begin{subfigure}[b]{0.3\textwidth}
         \centering
         \includegraphics[width=\textwidth]{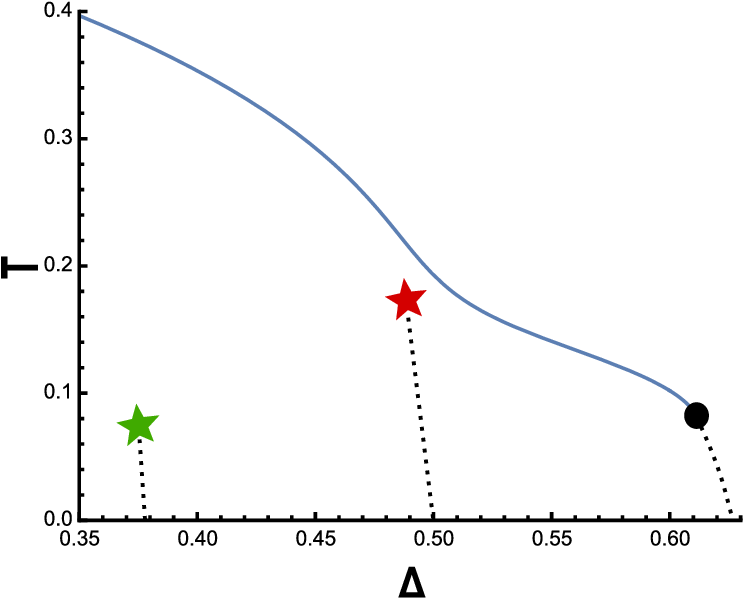}
        \caption{$ 0.4 \leq h < 0.577$}
         \label{fig63}
     \end{subfigure}
     \hfill
     \begin{subfigure}[b]{0.3\textwidth}
         \centering
         \includegraphics[width=\textwidth]{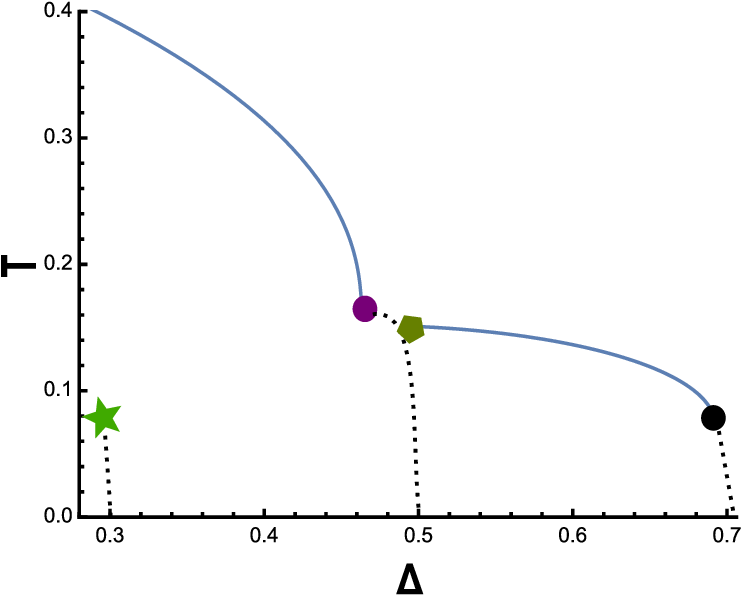}
        \caption{$ 0.577 \leq h \leq 0.6$}
         \label{fig64}
     \end{subfigure}
     \hfill
     \begin{subfigure}[b]{0.3\textwidth}
         \centering
         \includegraphics[width=\textwidth]{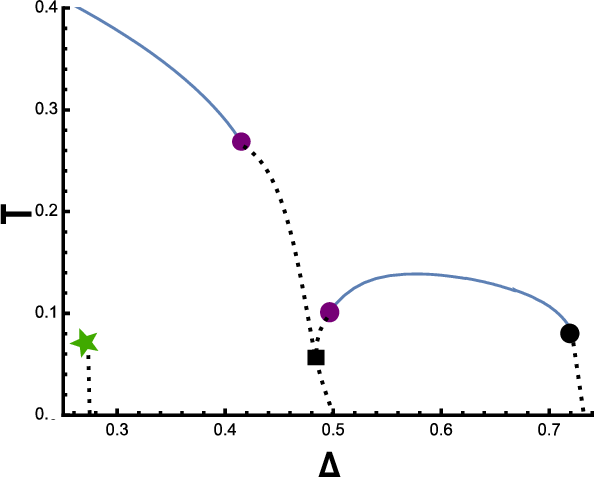}
        \caption{$ 0.6 \leq h < 0.625$}
         \label{fig65}
     \end{subfigure}
     \hfill
     \begin{subfigure}[b]{0.3\textwidth}
         \centering
         \includegraphics[width=\textwidth]{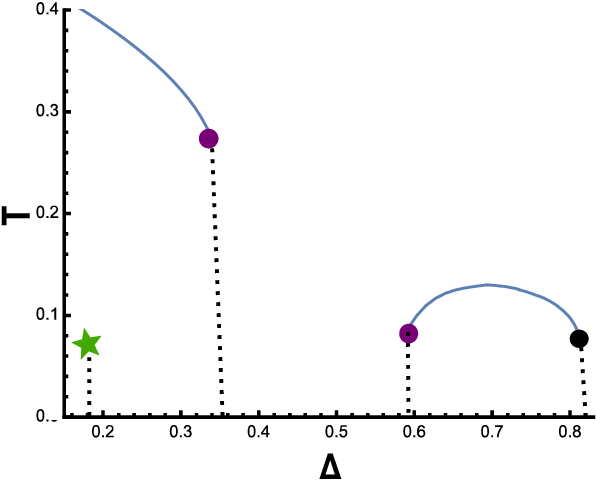}
        \caption{$ 0.625 \leq h < 0.875$}
         \label{fig66}
     \end{subfigure}
     \hfill
     \begin{subfigure}[b]{0.3\textwidth}
         \centering
         \includegraphics[width=\textwidth]{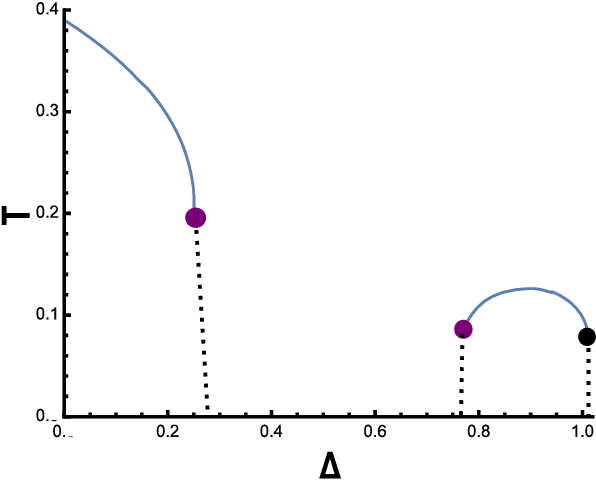}
        \caption{$  h > 0.875$}
         \label{fig67}
     \end{subfigure}
     \hfill
        \caption{ $T-\Delta$ phase diagram for different regimes of $h$ for $p=\frac{1}{2}$.  The solid line is the line of second order transition, the dotted lines are first order transitions, the solid stars are the BEPs,  solid circles are the TCPs,  black solid squares are the $A_5$ points and green circles are the CEPs.  There are eight different phase diagrams depending on the range of $h$.   }
        \label{fig6}
\end{figure}

For the $p=\frac{1}{3}$, the $T-h$ plane phase diagram shows re-entrance of the ordered phase for certain values of $\Delta$. For example, in Fig. \ref{fig102} and Fig. \ref{fig103} the phase diagram shows re-entrance. We study some thermodynamic quantities near the re-entrance regions (marked with  red cross in \ref{fig102} and \ref{fig103}). In Fig. \ref{fig999}, we plot of the magnetization ($m$), susceptibility ($\chi$) and specific heat ($C_v$) for  $h\, = 0.708$ and $\Delta=0.165$ (see Fig. \ref{fig102}). The $m$ shows a first order jump at low $T$ from \textbf{F3} to \textbf{P} then a small ordered region \textbf{F3} appears for higher $T$. Near the two continuous transition the $m$ can be fitted with the scaling function $ 2.008 \mid T-T_c \mid^{0.5}$ with $T_c= 0.131407$ and $0.628833 \mid T-T_c \mid^{0.5}$ with $T_c= 0.1615509$ respectively.  Both  continuous transitions lie in the mean-field Ising universality class.

\begin{figure}
     \centering
     \begin{subfigure}[b]{0.3\textwidth}
         \centering
         \includegraphics[width=\textwidth]{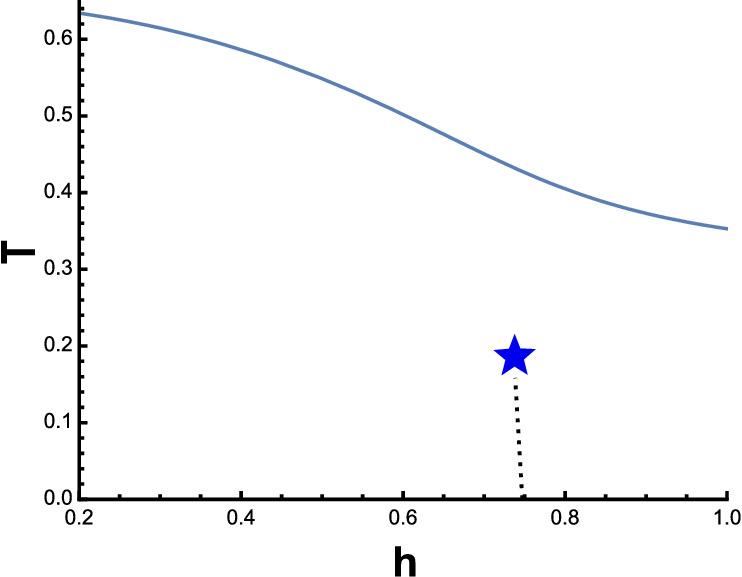}
         \caption{$ - \infty <  \Delta \leq 0.116$}
         \label{fig70}
     \end{subfigure}
     \hfill
     \begin{subfigure}[b]{0.3\textwidth}
         \centering
         \includegraphics[width=\textwidth]{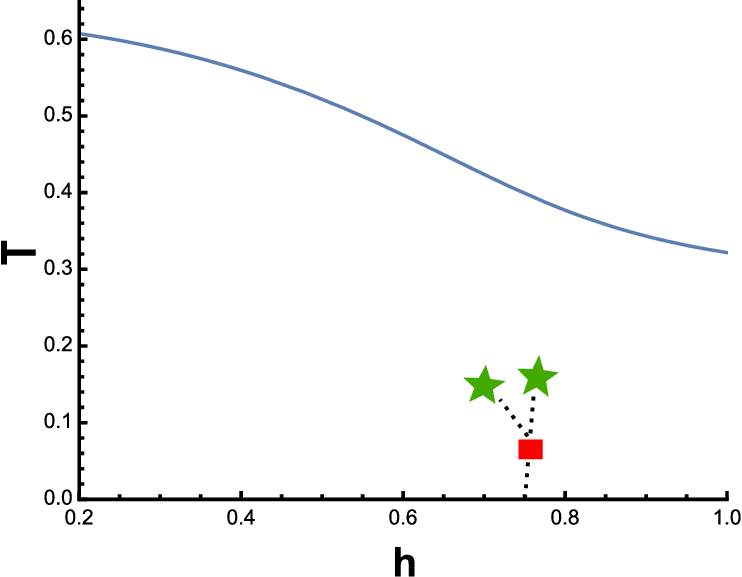}
          \caption{$ 0.116 < \Delta < 0.125$}
         \label{fig71}
     \end{subfigure}
     \hfill
     \begin{subfigure}[b]{0.3\textwidth}
         \centering
         \includegraphics[width=\textwidth]{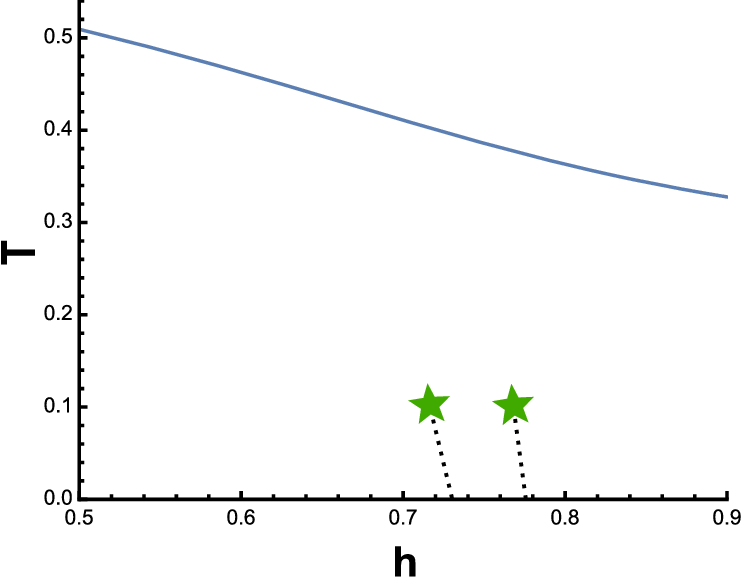}
        \caption{$ 0.125 \leq \Delta < 0.23$}
         \label{fig72}
     \end{subfigure}
     \hfill
     \begin{subfigure}[b]{0.3\textwidth}
         \centering
         \includegraphics[width=\textwidth]{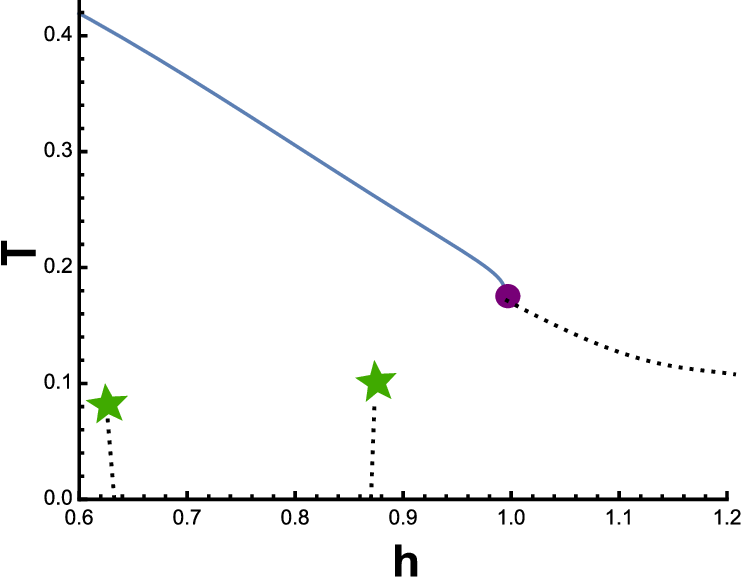}
        \caption{$ 0.23 \leq \Delta < 0.25$}
         \label{fig73}
     \end{subfigure}
     \hfill
     \begin{subfigure}[b]{0.3\textwidth}
         \centering
         \includegraphics[width=\textwidth]{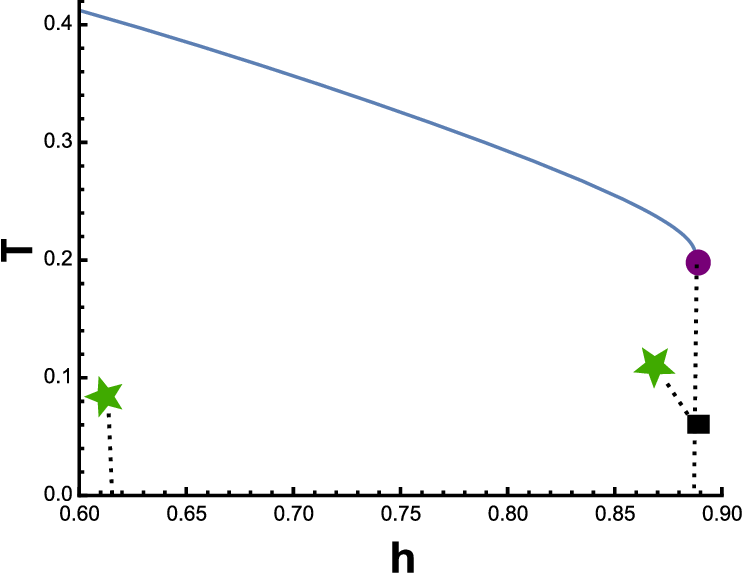}
        \caption{$ 0.25 \leq \Delta < 0.255$}
         \label{fig74}
     \end{subfigure}
     \hfill
     \begin{subfigure}[b]{0.3\textwidth}
         \centering
         \includegraphics[width=\textwidth]{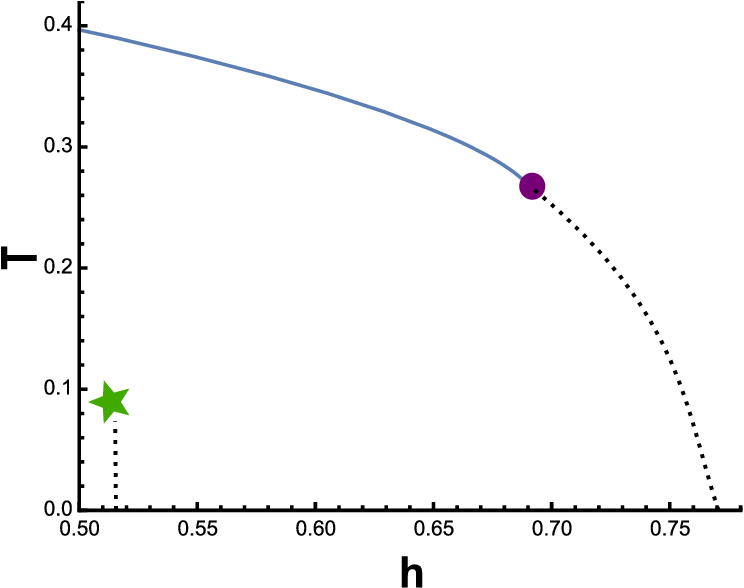}
        \caption{$ 0.255 \leq \Delta < \ln{4}/3$}
         \label{fig75}
     \end{subfigure}
     \hfill
     \begin{subfigure}[b]{0.3\textwidth}
         \centering
         \includegraphics[width=\textwidth]{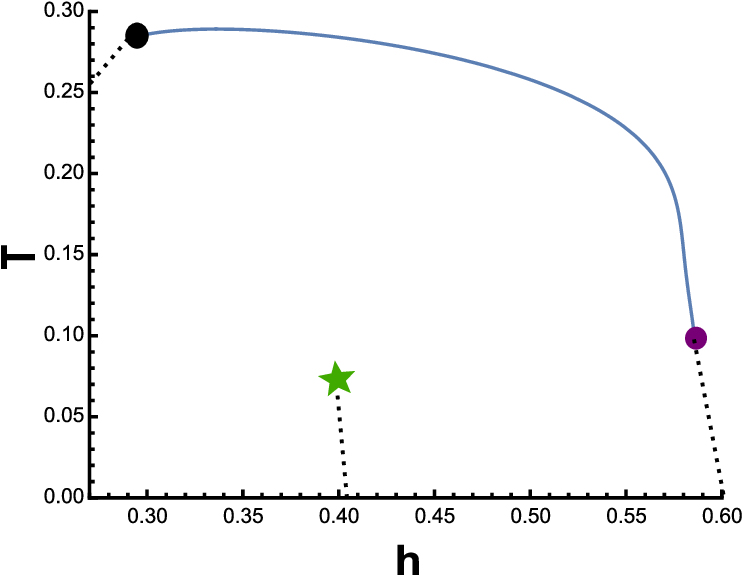}
        \caption{$ \ln{4}/3 \leq \Delta < 0.4988$}
         \label{fig751}
     \end{subfigure}
     \hfill
     \begin{subfigure}[b]{0.3\textwidth}
         \centering
         \includegraphics[width=\textwidth]{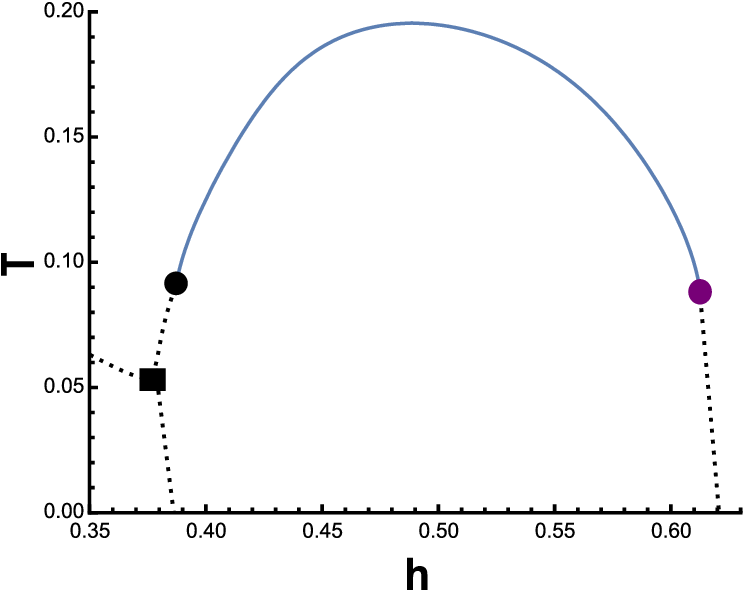}
        \caption{$ 0.4988 \leq \Delta < \frac{1}{2}$}
         \label{fig76}
     \end{subfigure}
     \hfill
     \begin{subfigure}[b]{0.3\textwidth}
         \centering
         \includegraphics[width=\textwidth]{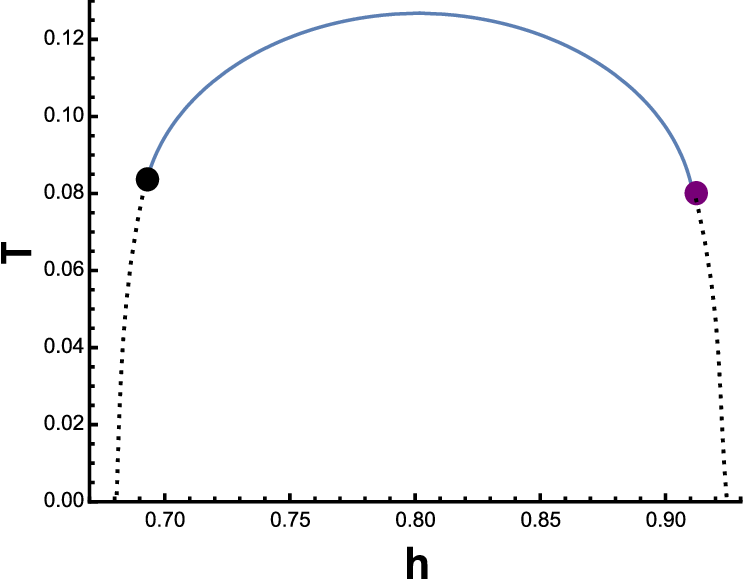}
        \caption{$  \Delta > \frac{1}{2}$}
         \label{fig77}
     \end{subfigure}
     \hfill
        \caption{ $T- h$ phase diagram for different regimes of $h$ for $p=\frac{1}{2}$.  The solid line is the line of second order transitions,  dotted lines are the lines of first order transitions,  solid stars are the BEPs,  solid circles are the TCPs,  black squares are the $A_5$ points  and red squares are the $A_6$ points.  There are nine different phase diagrams depending on the range of $\Delta$.   }
        \label{fig7}
\end{figure}

 In Fig. \ref{fig88} we plot  the magnetization($m$), susceptibility($\chi$)  and specific heat plot($C_v$) for two fixed values of $h\, = 0.692, \, 0.695$ at $\Delta=0.17$ (see  Fig. \ref{fig103}). The red and blue curves show the thermodynamic quantities for $h=0.692$ and $h=0.665$ respectively. For $h=0.692$ a small ordered region \textbf{F3} appears for higher $T$, showing re-entrance. Whereas for $h=0.665$, the magnetization ($m$) undergoes two transitions, a first order at low $T$ and a continuous transition at higher $T$ as shown in Fig. \ref{fig880}.   There is no re-entrance in the magnetization in this case.

\subsubsection{Phase diagrams for $p=\frac{1}{2}$}

  The Fig.\ref{fig6} shows the different phase diagrams in the $T-\Delta$ plane for different ranges of $h$ for $p=\frac{1}{2}$. There are eight different phase diagrams depending on the ranges of $h$. Three of them, Fig. \ref{fig60}, Fig. \ref{fig63} and Fig. \ref{fig67} are similar to the $T-\Delta$ phase diagrams of $p=\frac{1}{3}$ (Fig. \ref{fig90}, Fig. \ref{fig93} and Fig. \ref{fig98}). For $0.375 < h < 0.378$, there are three first order transition lines separating \textbf{F-F1}, \textbf{F1-F2} and \textbf{F2-NM} phases. The phase diagram contains one BEP, one TCP and one $A_5$ point (Fig.\ref{fig61}). As $h$ increases, the TCP  breaks into a CEP and a BEP. Thus there are two BEPs, one $A_5$ point and one CEP for $0.378 \leq  h < 0.4$  (see Fig.\ref{fig62}). For $0.4 \leq  h < 0.577$, the CEP and the $A_5$ point vanish (Fig.\ref{fig63}) and a TCP emerges. For $0.577 \leq  h \leq 0.6$, the second order transition line breaks into two parts and one BEP turns into a new TCP and a CEP (Fig.\ref{fig64}). For $0.6 \leq  h < 0.625$, the CEP breaks up into another TCP and a $A_5$ point and there are three TCPs, one BEP and one $A_5$ point in the phase diagram (Fig.\ref{fig65}).

In the $T-h$ plane projection (Fig.\ref{fig7}), there are nine different phase diagrams depending on the ranges of $\Delta$. Six of them (Fig. \ref{fig70} and Fig. \ref{fig74} - \ref{fig77}) are similar to the $T-h$ phase diagrams for $p=\frac{1}{3}$ (Fig. \ref{fig100} and Fig. \ref{fig103} - Fig. \ref{fig107}).  For $0.116 < \Delta < 0.125$, another first order transition line appears and the phase diagram consists of two BEPs and one $A_6$ point, see Fig.\ref{fig71}. For $0.23 \leq \Delta < 0.25$ one TCP emerges. Thus there are two BEPs and one TCP, see Fig.\ref{fig73}.

\subsubsection{Phase diagram for bimodal distribution ($p =0$)}

\begin{figure}
\centering
\includegraphics[scale=0.6]{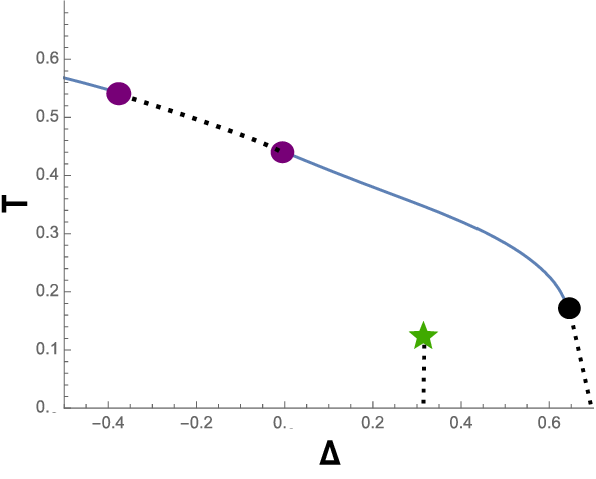}
\caption{$T-\Delta$ phase diagram for $0.414 \leq h < 0.4389$ for bimodal distribution ($p=0$). The solid line represents the line of continuous transition, the dotted line represents first order transition lines. The solid circles are the  TCPs and the solid star is  the BEP. The phase diagram has  three TCPs and one BEP. }
\label{fig-add}
\end{figure}

The bimodal distribution has been previously studied for the $T-h$ plane and the $T-\Delta$ plane in \cite{rfbc} and \cite{santos}respectively. We find  that, in the $T-h$ plane there are six  different  phase diagrams depending on the values of $\Delta$. These are similar to the $T-h$ phase diagrams  Fig. \ref{fig51} - \ref{fig56} for $p=\frac{1}{10}$.

Five  of these phase diagrams (Fig. \ref{fig51}, Fig. \ref{fig53}, \ref{fig54}, \ref{fig55} and \ref{fig56}) are similar as reported in \cite{rfbc} in the $T-h$ plane. In addition we find that for $0.244 < \Delta < 0.25$, there is a phase diagram similar to Fig. \ref{fig52}.

 In the $T-\Delta$ plane we find  that there are seven different types of phase diagrams depending on the ranges of $h$. Six of the phase diagrams are similar to the $T-h$ plane phase diagrams (similar to Fig. \ref{fig51} - Fig. \ref{fig56}). And the seventh phase diagram  for $0.414 \leq h < 0.4389$ consist of three TCPs and one BEP as shown in Fig. \ref{fig-add}.

In \cite{santos} the $T-\Delta$ plane phase diagram was reported for some distinct values of $h$. We re-obtain the five phase diagrams mentioned in \cite{santos}. We find two additional phase diagrams, one for $0.234 \leq h < 0.25$ similar to Fig. \ref{fig52} and another for $0.414 \leq h < 0.4389$ as shown in Fig. \ref{fig-add}. The phase diagram in Fig. \ref{fig-add} appears due to the non-monotonic behaviour of the locus of TCP. We will discuss this further in Sec. \ref{sec4}.

\subsection{Gaussian distribution}\label{sec3b}

Unlike the trimodal case,  the phase diagrams in $T-\Delta$ and $T-h$ plane for Gaussian distribution  contain only one TCP.

 Free energy functional for the Gaussian distribution at $H=0$ is,

\begin{eqnarray}\label{eq9}
f(m) = \frac{\beta m^2}{2}  - \frac{1}{\sqrt{2  \pi \sigma^2}}  \int_{-\infty}^{\infty}  \log (1+2 e^{-\beta \bigtriangleup} \cosh { \beta (m  +  h_i)}  )\,  e^{\frac{- h_i^2}{2 \sigma^2}}  dh_i 
\end{eqnarray}

Expanding Eq. \ref{eq9} around $m=0$ upto 8th order we get the following Landau coefficients

\begin{figure}
     \centering
     \begin{subfigure}[b]{0.45\textwidth}
         \centering
         \includegraphics[width=\textwidth]{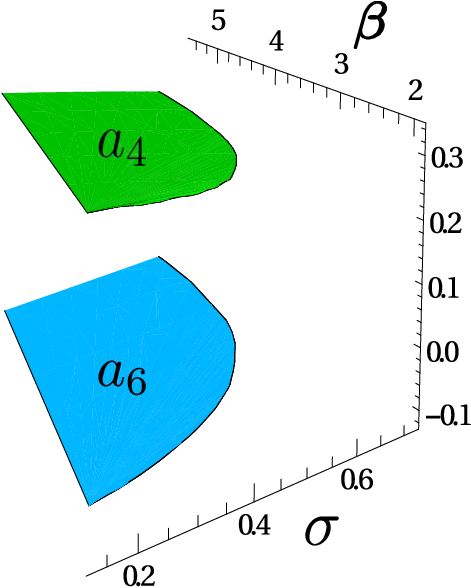}
         \caption{$   \Delta =0.3$}
         \label{fig200}
     \end{subfigure}
    \hfill
     \begin{subfigure}[b]{0.45\textwidth}
         \centering
         \includegraphics[width=\textwidth]{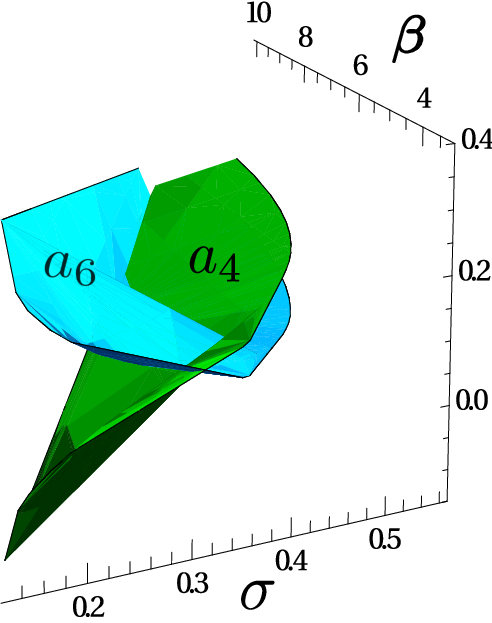}
          \caption{$  \Delta =0.465$}
         \label{fig201}
     \end{subfigure}
     \hfill
        \caption{ Plot of the Landau coefficients $a_4$ and $a_6$ when $a_2 =0$ in the $\beta - \sigma$ plane. Fig. \textbf{(a)} shows the plot for $\Delta=0.3$. Here $a_4$ is always positive at the coordinates of $a_2=0$. So the transition is always second order. Fig. \textbf{(b)} shows the plot for $\Delta=0.465$. For high values of $\sigma$, the $a_4$ is positive. So the transition is second order for this range of  $\sigma$. As $\sigma$ decreases, $a_4$ crosses $a_4=0$ at $\sigma_{th}=0.16$ and $\beta_{th}=3.2499$ provided $a_6 > 0$. So the second order transition ends at a TCP and becomes first order transition for $0 \leq \sigma < 0.16$.   }
        \label{fig20}
\end{figure} 

\begin{eqnarray}
a_2 &=& \frac{\beta}{2} -  \frac{ a \beta ^2}{\sqrt{2 \pi \sigma^2}}  \,\, \int_{-\infty}^{\infty} \frac{   z_1  +  2 a}{ \Big (1 + 2  a z_1 \Big )^2} \,\, e^{- \frac{h^2}{2 \sigma^2}}\,\, dh 
\end{eqnarray}

\begin{eqnarray}
a_4 =   \frac{a \beta^4 }{ 12 \sqrt{2 \pi \sigma^2}}  \,\, \int_{-\infty}^{\infty} \frac{   4 a ( 4 a^2 -1)(1-z_2) - a^2 (z_3 - 16 a) +(  13 a^2 -1 ) z_1  }{  \Big (1 + 2 a z_1 \Big )^4}  \,\, e^{- \frac{h^2}{2 \sigma^2}} dh \nonumber \\
\end{eqnarray}

\begin{eqnarray}
a_6 = -  \frac{ a \beta^6}{360 \sqrt{2 \pi \sigma^2}}   \int_{-\infty}^{\infty} && \Bigg ( 6 a(352 a^4 - 69 a^2 +1) + 
   (898 a^4 - 146  a^2+  1) z_1 
   -
  26 a(64 a^4 - 20  a^2 +  1) z_2 - 3 a^2 ( 
  113  a^2  - 33 ) z_3 \nonumber \\
  &+& 
  2 a^3 (32 a^2 - 13  ) z_4 +  a^4 z_5 \Bigg ) \Big (1 + 2 a z_1 \Big )^{ -6}  \,\, e^{- \frac{h^2}{2 \sigma^2}}\,\, dh 
\end{eqnarray} 
where  $a= e^{-\beta \Delta}$ and $z_n= \cosh { n \beta h}$.

\begin{figure}
     \centering
         \includegraphics[scale=0.7]{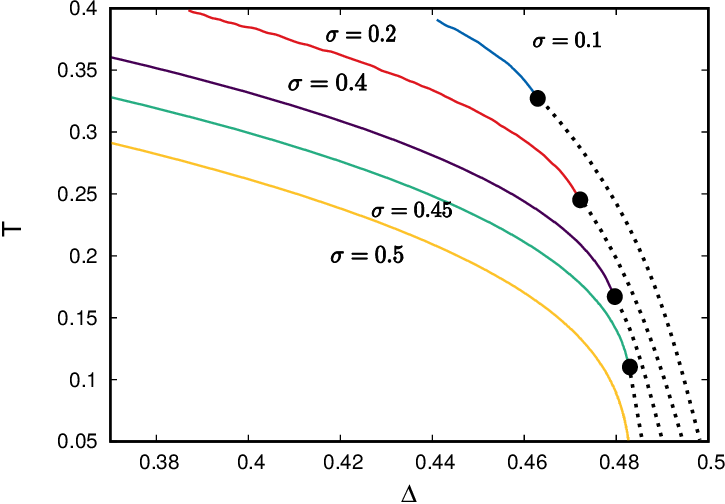}
         \caption{Phase diagrams of the Gaussian RFBC in the $T-\Delta$ plane for different values of $\sigma$. The solid lines are the lines of second order transition and the dotted lines are the first order transition lines. Solid circles are the TCPs. For $\sigma \,\, < \, \sigma_{TCP} \approx 0.4839$ the phase diagram exhibits a TCP. This TCP moves to $T=0$ at $\sigma_{TCP}$. For $\sigma \, > \, \sigma_{TCP}$ there is only a line of continuous transition in the phase diagram.}
         \label{fig131}
 \end{figure}

Integrating $a_2$ numerically and then equating it to zero, we get the line of continuous transition, provided that $a_4>0$ at those coordinates. Similarly, integrating $a_4$ at the coordinates of $a_2=0$ and then equating it to zero gives the location of the TCP, provided $a_6>0$. In order to obtain the coordinates of the $\lambda$ line and the TCP in the $T-\sigma$ plane and $T-\Delta$ plane  we plot the values of the $a_4$ and $a_6$ coefficients after substituting the coordinates for which $a_2=0$ for different values of $\Delta$ and $\sigma$ respectively.

To illustrate the procedure, in Fig. \ref{fig20} we plot the $a_4$ and $a_6$ values for the condition $a_2=0$  at two values of $\Delta$. Fig. \ref{fig200} shows the plot of $a_4$ and $a_6$ for fixed $\Delta =0.3$. In this case $a_4>0$ and the transition is always second order. Fig. \ref{fig201} shows the plot for $\Delta=0.465$. Here we find  $a_4 >0$ for $ \sigma > 0.16$. At $\sigma_{th}=0.16$ and $\beta_{th} = 3.2499$,  $a_4 =0$ with $a_6 > 0$. This  hence is the locus of the TCP. For $\sigma < \sigma_{th} $ the  transition is always first order. The coordinates of the first order transition can be found by equating the free energies ($f(m=0) = f(m\neq 0)$) and also their first order derivative on both side. We use this method to obtain the phase diagram in the entire $T-\sigma$ and $T-\Delta$ planes by fixing the values of $\Delta$ and $\sigma$ respectively.

     \begin{figure}
          \centering
         \includegraphics[scale=0.7]{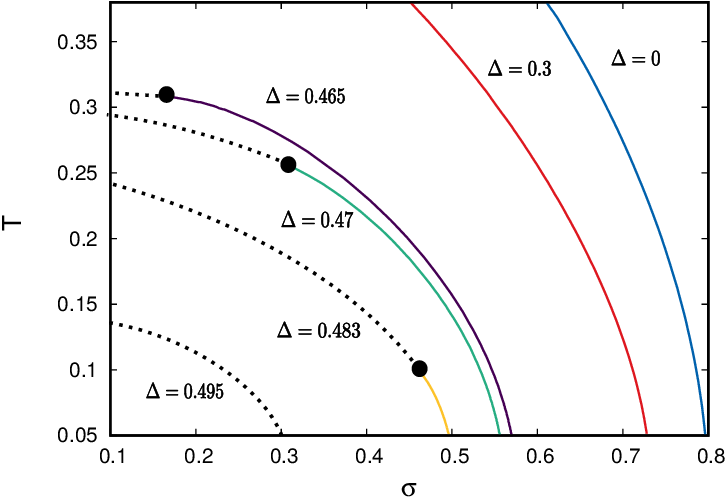}
          \caption{Phase diagrams of the Gaussian RFBC in the $T-\sigma$ plane for different values of $\Delta$. Below $\Delta = \Delta_{BC} = \frac{\ln 4}{3} \approx 0.462098$, there is only a line of  second order transition. For $\Delta> \Delta_{BC}$, TCP emerges and moves to $T=0$ at $\Delta_{TCP}=\sigma_{TCP}\simeq 0.4839$. There are only first order transition lines for $  0.5\geq \Delta> \Delta_{TCP}$. For $\Delta > 0.5$, there is no ordered state and hence no transition. }
         \label{fig132}
\end{figure}

 We find that for small values of $\sigma$, the transition in the $T-\Delta$ plane is second order at high temperature and first order at low temperature. These two transition lines meet at a TCP. As  $\sigma$  increases, the first order transition line decreases and above $ \sigma_{c}= \sqrt{\frac{2}{e \pi}} \sim 0.483941..$, the transition becomes second order. This is same as the value of $\sigma_{TCP}$ in Sec. \ref{sec2b}. There $\sigma_{TCP}$ was the TCP value of $\sigma$ for $T=0$, below which the transition is always first order. For $\sigma < \sigma_{c}$, there is always a TCP in the $T-\Delta$ phase diagram shown in Fig.  \ref{fig131}.

  Similarly in the $T-\sigma$ plane, the phase diagram consists of a second order transition for $\Delta < \Delta_{BC} (= \frac{\ln 4}{3})$, which is the value of $\Delta$ at the TCP of pure BC model. For $\Delta \geq  \Delta_{BC}$, one TCP emerges in the phase diagram and the phase diagram consists of first and a second order transition lines meeting at a TCP. The TCP moves to lower temperature with the increasing $\Delta$. At exactly $\Delta= \Delta_{TCP}= \sqrt{\frac{2}{e \pi}} \sim 0.483941..$ (TCP value at $T=0$), the TCP moves to zero and there is only a first order transition line in the phase diagram (shown in Fig.\ref{fig132}). For $  \Delta_{TCP} < \Delta \leq \frac{1}{2}$, the transition is always first order and for $\Delta > \frac{1}{2}$, there is no transition in the $T-\sigma$ plane.

 In a recent study of spin-$s$ random field Blume Capel model using the Gaussian distribution using effective field theory only continuous transition lines were reported \cite{sspin}. They did not find the lines of first order transition and the TCP.

\section{Multicritical points in the phase diagrams of the trimodal distribution}\label{sec4}

\begin{figure}
     \centering
       \begin{subfigure}[b]{0.4\textwidth}
         \centering
         \includegraphics[width=\textwidth]{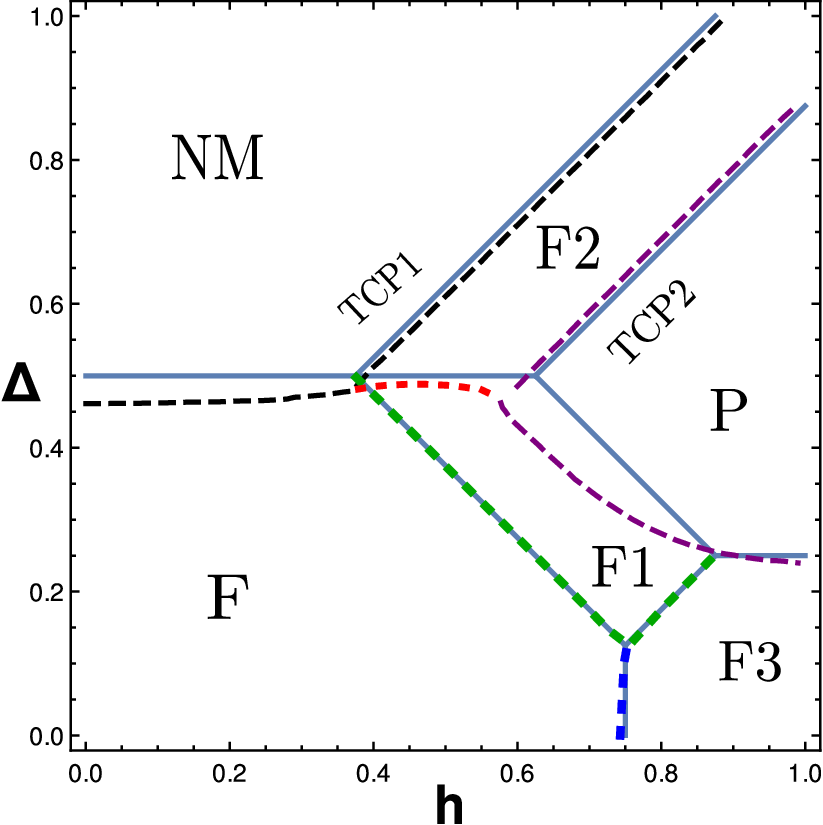}
        \caption{$p = \frac{1}{2}$}
         \label{fig142}
     \end{subfigure}
     \hfill
      \begin{subfigure}[b]{0.4\textwidth}
         \centering
         \includegraphics[width=\textwidth]{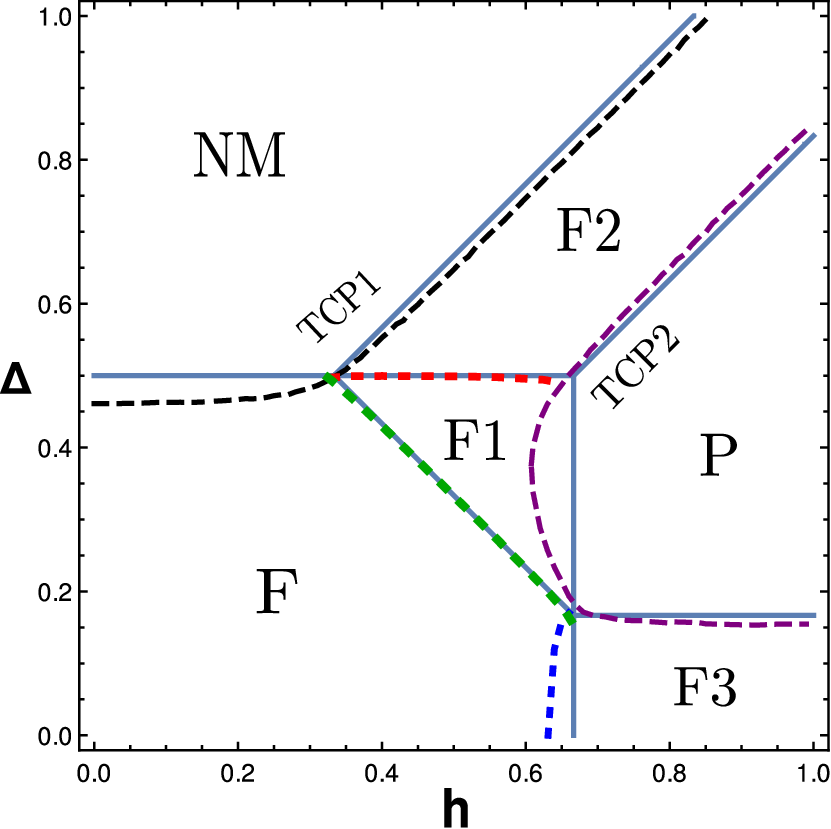}
          \caption{$p = \frac{1}{3}$}
         \label{fig141}
     \end{subfigure}
     \hfill
     \begin{subfigure}[b]{0.4\textwidth}
         \centering
         \includegraphics[width=\textwidth]{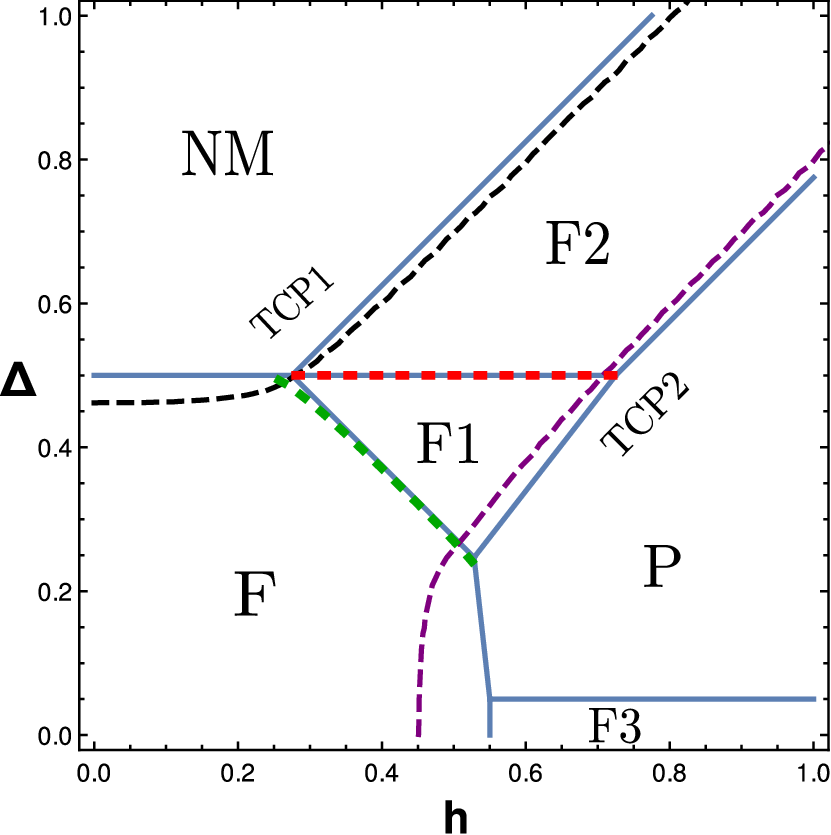}
         \caption{$p = \frac{1}{10}$}
         \label{fig140}
     \end{subfigure}
  \hfill
  \begin{subfigure}[b]{0.4\textwidth}
         \centering
         \includegraphics[width=\textwidth]{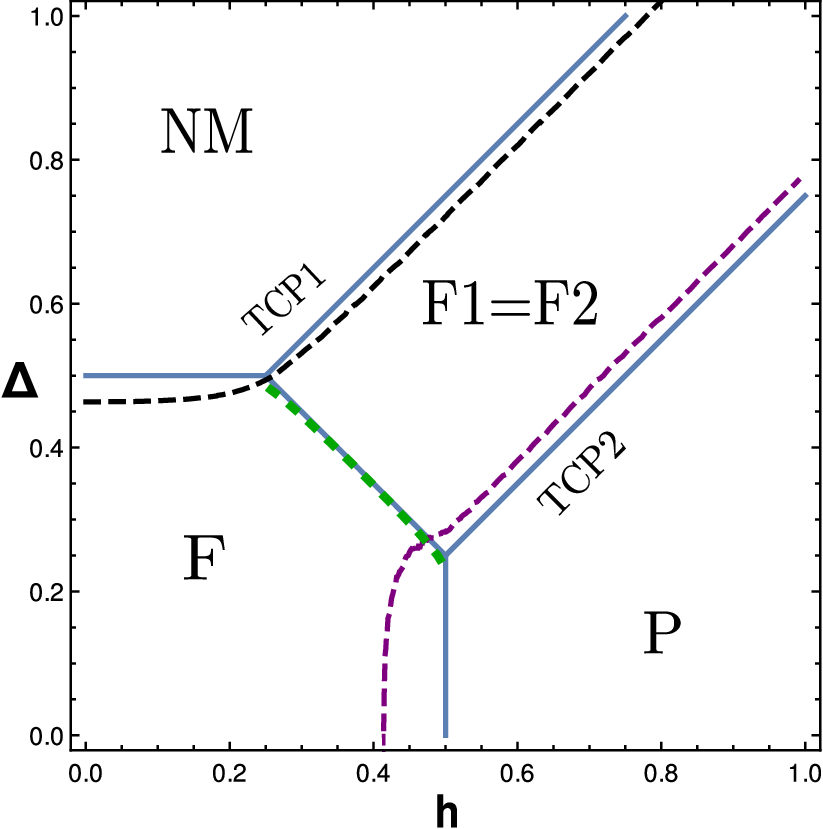}
         \caption{$p = 0$}
         \label{fig144}
     \end{subfigure}
  \hfill
        \caption{Projection of the TCP and BEP coordinates at different $T$ onto the ground state phase diagram for  \textbf{(a)} $p=\frac{1}{2}$, \textbf{(b)} $p=\frac{1}{3}$, \textbf{(c)} $p=\frac{1}{10}$, and \textbf{(d)} $p=0$.   The solid blue lines are the ground state phase boundaries,  the black dashed lines are the projection of the TCP1 coordinates, the purple dashed lines are the projection of the TCP2 coordinates, and the red, green and blue dashed lines are the projection of the BEP coordinates along the phase boundaries of the different phases.  The TCP2 coordinate is non-monotonic depending on the value of $p$. }
        \label{fig14}
\end{figure}

For the trimodal random field distribution, the RFBC model exhibits six phases,  multicritical points like BEP,  CEP and multi-phase coexistence points like  $A_5$,  $A_6$ and $A_7$ along with the TCPs.

We projected the coordinates of TCPs and BEPs on the ground state phase diagram in the $\Delta-h$ plane (Fig. \ref{fig14}). The solid blue line shows the phase boundaries in the ground state,  black and purple dashed lines  are the projections of the coordinates of the TCPs and red, blue and green dotted lines are the projections of the coordinates  of the BEPs. We have not shown the pure case ($p=1$) as  there is only one TCP  which appears at $T=\frac{1}{3},  \, \, \Delta=0.462098,  \,\,  h=0$.  As we switch on disorder by taking $p$ less than $1$,  the coordinate of this TCP (we call it as TCP1) increases monotonically in  $\Delta$ as $h$ increases (shown by black dashed lines). Fig. \ref{fig142} is the plot for the projection of the TCP and BEP coordinates for $p=\frac{1}{2}$.  Along with this TCP1 line,  another line of TCP emerges (shown by purple dashed lines) along the phase boundaries of  \textbf{F2 - P} , \textbf{F1 - P } and \textbf{F3 - P } (we call it as TCP2).  Along with the new TCP2 line, three  BEP lines also emerge  along the separation of the phases \textbf{F-F3},  \textbf{F-F2}, \textbf{F-F1}, \textbf{F1-F3} denoted by blue, red and green dotted lines respectively.

  On further decreasing  $p$,  the BEP line along the phase separations of \textbf{F3-F1} vanishes and there are now three BEP lines along the phase separation lines of \textbf{F1-F2} and \textbf{F-F3} and \textbf{F-F1}.  The TCP1 behaves similarly to the $p=\frac{1}{2}$ case.  And the TCP2 starts from ($\Delta \simeq \frac{p}{2},  \,\,  h \rightarrow \infty$). As $h$ decreases, the TCP2 line remains close to $\Delta \simeq \frac{p}{2}$ until $h \simeq \frac{1+p}{2}$. Below $h \simeq \frac{1+p}{2}$, the TCP2 line shows an extrema at $\Delta \approx 0.375$, $h \approx 0.607$, $T \approx 0.24$ and then increases in $\Delta$ as $h$ increases. Due to this extrema the TCP2 line shows non-monotonic behaviour. Fig. \ref{fig141} shows the projection of the BEPs and TCPs for $p=\frac{1}{3}$.

  For $p$ moving towards the bimodal value $p=0$,  the line of BEPs  along the \textbf{F-F3} phase separation vanishes and the phase diagram now consist of two lines of BEP  along the phase boundaries of  \textbf{F1-F2} and \textbf{F-F1}. The TCP1 behaves similar to as for $p=\frac{1}{3}$. The TCP2 line  starts from $\Delta \rightarrow - \infty$ instead of $\Delta \simeq \frac{p}{2}$ and then $\Delta$ increases as $h$ increases (shown in Fig. \ref{fig140} for $p=\frac{1}{10}$).

  \begin{figure}
     \centering
       \begin{subfigure}[b]{0.32\textwidth}
         \centering
         \includegraphics[width=\textwidth]{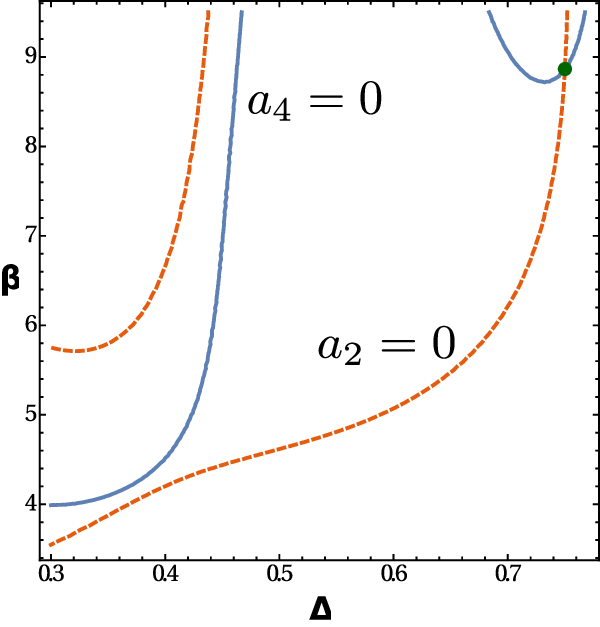}
        \caption{$h = 0.603 $}
         \label{fig0650}
     \end{subfigure}
     \hfill
      \begin{subfigure}[b]{0.32\textwidth}
         \centering
         \includegraphics[width=\textwidth]{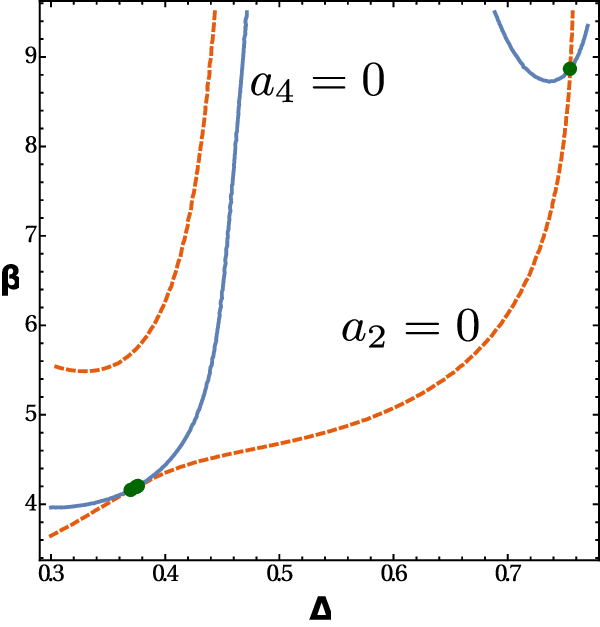}
          \caption{$h= h_c = 0.6079$}
         \label{fig0651}
     \end{subfigure}
     \hfill
     \begin{subfigure}[b]{0.32\textwidth}
         \centering
         \includegraphics[width=\textwidth]{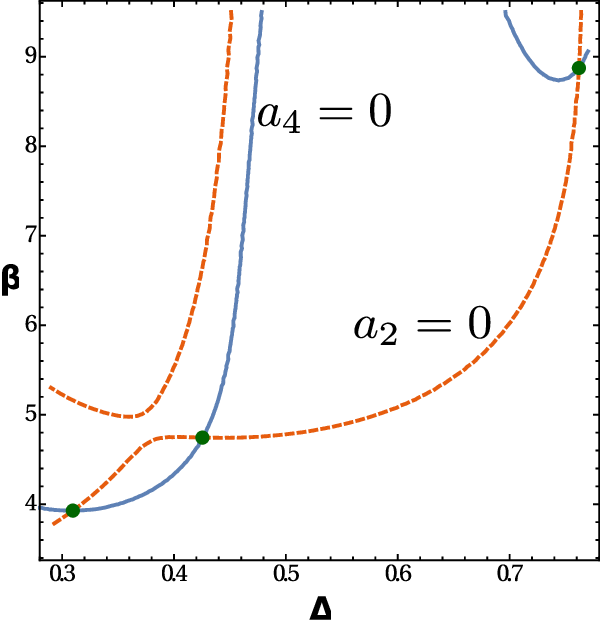}
         \caption{$h = 0.615 $}
         \label{fig0652}
     \end{subfigure}
  \hfill
        \caption{Contour plot of  the $\lambda$ line (Eq. \ref{eq4}) shown by dashed lines and the solutions for $a_4=0$ given by Eq. \ref{a4=0} (solid blue line) for $p=\frac{1}{3}$ in the $\beta-\Delta$ plane for $h$ close to $h_c$. Here $h_c$ is the value of the random field at which the double TCP emerges. \textbf{(a)} For $h = 0.603$, the two curves intersect only once (shown by solid green circle), hence the phase diagram Fig. \ref{fig93} shows only one TCP (which is TCP1). \textbf{(b)} For $h = h_c = 0.6079$, the two curves intersect at TCP1 for higher $\Delta$ and for low $\Delta$ the two curves are tangential to each other.  \textbf{(c)} For  $h = 0.615$, the two curves intersect at three points giving rise to one TCP1 and two TCP2s in the phase diagram Fig. \ref{fig95}.  }
        \label{fig065}
\end{figure}

   At exactly $p=0$,  we get back the two TCP lines (TCP1 and TCP2) of the bimodal distribution (shown in Fig. \ref{fig144}). TCP1 starts from the pure Blume-Capel model TCP and then increases monotonically with increasing $\Delta$ and $h$. The TCP2 starts at the TCP of the RFIM  $\Delta\rightarrow - \infty,  \,\,  T=\frac{2}{3},  \,\,  h\sim 0.43899$ and increases non-monotonically in $\Delta$ as $h$ increases. The phase diagram also exhibits one BEP line along the phase separation of \textbf{F-(F1=F2)}.

  By studying the projection of TCPs and BEPs on the ground state phase diagram, we find that their coordinates closely follow the phase boundaries present in the $T=0$ phase diagrams. We hence show that the  multicritical points arise due to the presence of first order transition lines in $T=0$ phase diagram.

   The TCP2 line shows a non-monotonic dependence of $\Delta$ on $H$ for all $0 \leq p < 1$. As a result as we cross the TCP2 line along $\Delta$ axis near the non-monotonic regime, we get three TCPs in the phase diagrams of $T-\Delta$ plane for some values of $h$ (i.e Fig. \ref{fig43} for $p = \frac{1}{10}$, Fig. \ref{fig95} and Fig. \ref{fig99} for $p=\frac{1}{3}$, Fig. \ref{fig65} for $p=\frac{1}{2}$ and Fig. \ref{fig-add} for $p=0$). Two intercepts come from the TCP2 line and the other comes from the TCP1 line. These two TCP2 points emerge in the phase diagram as a pair.  For example in Fig. \ref{fig065} we plot the values of $(\beta, \Delta)$ at which the second order term ($a_2$) equals zero. These are the coordinates of the $\lambda$ line (Eq. \ref{eq4}). We also plot the values of $(\beta, \Delta)$ at which the third order term ($a_4$) equals zero in the expansion of Eq. \ref{eq2}. The equation for  $a_4=0$ is 
   
 \begin{eqnarray}\label{a4=0}
 \frac{\beta^4 a }{12} \Bigg (  \frac{p(4 a -1)}{(2 a +1)^2} - \frac{(1-p) (a^2 (z_3 - 13) -16 a^3 (2+ z_1) + 4 a (1 - z_2) +z_1)}{(1+ 2 a z_1)^4}   \Bigg ) =0 \nonumber \\
 \end{eqnarray}
   
  We plot the solutions of Eq. \ref{eq4} (shown by red dashed line) and Eq. \ref{a4=0} (shown by solid blue line)  in the $\beta-\Delta$ plane for $h$ values very close to the value $h_c$ at which two TCP2 emerge for $p=\frac{1}{3}$. Fig. \ref{fig0650} shows for $h < h_c$, the two curves intersect only at one point and the  phase diagram (Fig. \ref{fig93}) consists of only TCP1. Near $h \approx h_c$, the two curves almost become tangential, see Fig. \ref{fig0651}. For $h \geq h_c$ the curves intersect thrice giving rise to two TCP2s and one TCP1 in the phase diagram (Fig. \ref{fig95}), see Fig. \ref{fig0652}.

 \section{Discussion} \label{sec6}

We studied the RFBC model with trimodal distribution of the random field on a fully connected graph that has not been studied earlier. We find many different phase diagrams depending on the values of $p$, $\Delta$ and $h$. One striking feature of these phase diagrams is the presence of many multicritical and multi-coexistence points. Depending on the values of $p$, $\Delta$ and $h$, the model exhibits multiple first order transitions as a function of the  temperature. Reentrance was also seen for a narrow range of parameters for $p=\frac{1}{3}$. For $p=0$, the bimodal distribution, besides obtaining the phase diagrams reported earlier \cite{rfbc, santos}, we also obtain two new phase diagrams. 

The RFBC model in the presence of trimodal distribution shows re-entrance at low temperatures for $p=\frac{1}{3}$ in the $T-h$ phase diagram for a range of $\Delta$. The mean-field Blume-Capel model are known to show re-entrance in the presence of strong degeneracy of the $s= \pm 1$ states \cite{reentrance, reentrance1}. In the RFBC, in some region of the parameters, the energy gain due to $s=\pm 1$ spin is unable to compensate the entropy loss and the system chooses to increase its entropy by increasing the density of $s=0$ spins and the ordered state is lost as the temperature is reduced as shown in Fig. \ref{fig990} and Fig. \ref{fig880}.

For the Gaussian distribution of the random field, we found much simpler phase diagrams with either one TCP or none. This differs from the earlier study using effective field theory (EFT) \cite{sspin} which reported only continuous transitions for all strengths of the Gaussian random field.

It was shown that for $\frac{1}{3} \leq p <1$ the RFIM shows similar phase diagram  for Gaussian and symmetric trimodal distributions \cite{trimodal1, trimodal2, trimodal3, trimodal4, numerical2}.  We find that for RFBC model the phase diagram for the Gaussian random field is different from that of the symmetric trimodal distribution. The argument was based on $\epsilon$ expansion study of the random field $O(n)$ models \cite{aharony} and hence need not hold for higher spin models like RFBC.  Another possibility is that the difference can  be an artefact of mean-field nature of the calculations.  It would be useful to study different symmetric random field distributions for the RFBC model using simulations in the finite dimensions \cite{numerical}  to check if such differences would still exists in finite dimensions.

The information of the phases at $T = 0$ was found to be useful in understanding the behaviour at finite $T$.  For trimodal distribution, the value of $\Delta$ and $h$ at which the multicritical points like TCPs and BEPs appear at finite temperatures were observed to be close to the first order transition lines in the ground state $\Delta - h$ phase diagram. This suggests that the random field dominates the low temperature behaviour. The interplay of $\Delta$ and $h$ results in many stable phases in the ground state.  These phases have different configurational entropy.  This plays a crucial role in determining the finite temperature  behaviour of the system.

In the case of trimodal distribution at finite $T$, we observed two TCPs connected via  a first order transition line in the $T-\Delta$ plane. The locus of these TCPs was found to come closer on changing $h$ and disappear eventually. This kind of behaviour was recently seen in non-equilibrium transitions in the resetting problem \cite{dibyendu}. To understand their origin, it would be interesting to look at the first order wings originating from these TCPS and BEPs by applying a uniform external field as was done in the case of bimodal random crystal field Blume-Capel model \cite{sumedha}.

Another interesting model is the Blume-Capel model in the presence of random crystal field (RCFBC). This model has been studied  earlier for the bimodal \cite{sumedha} as well as for the Gaussian distributions \cite{gausscrystal}. The effect of the random field disorder is different than the effect of random crystal field disorder as can be seen in the nature of the phase diagrams. A study of trimodal distribution for RCFBC has not been done. The method used in this paper can be straightforwardly applied to many other models like Blume-Emery-Griffiths model (BEG). The BEG model has been studied in  the presence of random crystal field \cite{branco}. It would be interesting to look at the effect of random field on BEG model.

\appendix
\renewcommand{\theequation}{A-\arabic{equation}}
\setcounter{equation}{0}
\renewcommand{\thefigure}{A-\arabic{figure}}
\setcounter{figure}{0}

\section{Rate function for the RFBC}\label{sec1b}

The probability of a spin configuration $C_N$ with magnetization $x_1 = \frac{\sum \limits_i s_i}{N}$ and quadrupole moment $x_2 = \frac{\sum \limits_i s_i^2}{N}$ is proportional to $ e^{- \beta \mathcal{H}}$, where $\mathcal{H}$ is the Hamiltonian given in Eq. \ref{eq1}. This via large deviation principle (LDP) in the limit of $N \rightarrow \infty$  goes to $P(C_N) \sim e^{- N I(x_1, x_2)}$.  The function $I(x_1, x_2)$ here is the rate function which is like the generalized free energy functional. To calculate $I(x_1, x_2)$ we use two steps :

\begin{enumerate}

\item  Calculate the rate function $R(x_1, x_2)$ corresponding to the probability $P_{\mathcal{H}_{ni}} (C_N) \sim e^{- N R(x_1, x_2)}$. Here $\mathcal{H}_{ni}$ is the non-interacting part of the  Hamiltonian i.e $\mathcal{H}_{ni} = \Delta \sum_i s_i ^2 - \sum_i (h_i+H) s_i$.
The function $R(x_1, x_2)$ is calculated using the G\"{a}rtner-Ellis (GE) theorem \cite{ldp}. GE theorem states that $R(x_1, x_2)$ is given by the Legendre-Fenchel transformation of the scaled cumulant generating function $\lambda(k_1, k_2)$, provided $\lambda(k_1, k_2)$ is differentiable. The expression of $R(x_1, x_2)$ is 
\begin{eqnarray}\label{eqr}
		R(x_1,x_2) &=&  \sup_{k_1,k_2} \Bigg [x_1k_1+x_2k_2 - \lambda(k_1, k_2) \Bigg ] \nonumber \\
	\end{eqnarray}
	
The function $\lambda(k_1, k_2) = \lim \limits_{N \rightarrow \infty} \frac{1}{N} \lambda_N(k_1, k_2)$ where $\lambda_N (k_1, k_2)$ is the logarithmic cumulant generating function of $x_1$ and $x_2$ w.r.t the probability $P_{\mathcal{H}_{ni}}$. The  $\lambda(k_1, k_2)$ for the random variables $x_1$ and $x_2$ is given by

\begin{eqnarray}
\lambda(k_1, k_2) =
\Bigg < \log (1+2 e^{k_2-\beta \bigtriangleup} \cosh (k_1+ \beta H+ \beta h_i))\Bigg > \nonumber \\
\end{eqnarray}
$\langle \rangle$ represents the average over the random field distribution.

 Minimization of the expression $x_1k_1+x_2k_2 - \lambda(k_1, k_2)$ in Eq. \ref{eqr} w.r.t $k_1$ and $k_2$ gives the following equations for the supremum  ($k_1^*$, $k_2^*$) as a function of $x_1$ and $x_2$

\begin{eqnarray}\label{eqn30}
		x_1 =  \Bigg< \frac{ 2 e^{k_2^* -\beta \Delta} \sinh (\beta h_i+ \beta H +k_1^*)}{1+ 2 e^{k_2^* -\beta \Delta} \cosh (\beta h_i+ \beta H + k_1^*)}\Bigg > \nonumber \\
	\end{eqnarray}
\begin{eqnarray}\label{eqn20}
		x_2 =  \Bigg< \frac{2 e^{k_2^* -\beta \Delta} \cosh (\beta h_i+ \beta H +k_1^*)}{1+ 2 e^{k_2^* -\beta \Delta} \cosh (\beta h_i+ \beta H + k_1^*)}\Bigg > \nonumber \\ 
	\end{eqnarray}

\item The full rate function of the interacting Hamiltonian can be calculated via tilted LDP \cite{hollander}. This principle allow us to calculate the  rate function $I(x_1, x_2)$ from the old rate function ($R(x_1, x_2)$) using a change in measure by integrating against an exponential of a continuous function $G(x_1, x_2)$ which in our case is  the interacting part of the Hamiltonian, $ G = \frac{\beta x_1^2}{2}$. The  rate function $I(x_1, x_2)$ is given by (see \cite{disc-ldt, cont-ldt} for more details)

\begin{equation}
I(x_1, x_2) = R(x_1, x_2) - \frac{\beta x_1^2}{2} - \inf \limits_{y_1, y_2} \Big (R(y_1, y_2) - \frac{\beta y_1^2}{2} \Big )
\end{equation}

After substituting $R(x_1, x_2)$ we get
	
	\begin{eqnarray}\label{eqI}
		I(x_1, x_2) & = & x_1 k_1^* + x_2 k_2^* - \frac{\beta x_1^2}{2}   -  \,\, \Bigg < \log (1+2 e^{k_2^*-\beta \bigtriangleup} \cosh (k_1^*+ \beta H+ \beta h_i))\Bigg > \nonumber \\
	\end{eqnarray}
	here ($k_1^*, \,\, k_2^*$) are given by the solutions of Eqs. \ref{eqn30} and \ref{eqn20}. Minimizing the full rate-function w.r.t the order parameters ($x_1$, $x_2$) we get $k_1^*= \beta m$ and $k_2^*=0$. The variables $m$ and $q$ represent the minimum of $x_1$ and $x_2$ respectively. On substituting $k_1^*$ and $k_2^*$ in Eq. \ref{eqI} we get the  free energy functional to be

\begin{eqnarray}
		f(m) =  \frac{\beta m^2}{2}- \,\, \Bigg < \log (1+2 e^{-\beta \bigtriangleup} \cosh { \beta (m + H +  h_i)} \,\, )\Bigg > \nonumber \\
	\end{eqnarray}

\end{enumerate}

\end{document}